\documentclass[review,authoryear]{elsarticle}
\usepackage[utf8]{inputenc}
\usepackage[T1]{fontenc}
\usepackage{kpfonts}
\usepackage{textcomp}

\usepackage{graphicx}
\usepackage[english]{babel}
\usepackage[table]{xcolor}
\usepackage{amsmath,amsfonts,amssymb}
\usepackage[version=3]{mhchem}
\usepackage{array}

\usepackage{fullpage}
\usepackage{subfig}
\usepackage{booktabs}								
\usepackage{multirow}	
\usepackage[section]{placeins}
\usepackage{tabularx}
\usepackage{natbib}
\usepackage{fancyref}

\usepackage{lineno}

\newcommand{\Da}{Damk\"ohler}

\begin{document}

\title{Reactive transport experiments of coupled carbonation and serpentinization in a natural serpentinite. Implication for hydrogen production and carbon geological storage}
\author[1]{F. Osselin\corref{cor1}}
\ead{florian.osselin@cnrs-orleans.fr}
\author[1]{M. Pichavant}
\author[1]{R. Champallier}
\author[2]{M. Ulrich}
\author[1]{H. Raimbourg}

\cortext[cor1]{Corresponding author}
\affiliation[1]{organization={Institut des Sciences de la Terre d'Orléans, Université d'Orléans/CNRS/BRGM UMR7327}, 
addressline={1A Rue de la Ferollerie}, 
postcode={45100}, 
city={Orléans}, 
country={France}}

\affiliation[2]{organization={Université de Strasbourg, CNRS, Institut Terre et Environnement de Strasbourg, UMR7063},
addressline={5 Rue Descartes}, 
city={Strasbourg},
postcode={F-67084},
country={France}}

\maketitle 

\linenumbers

\section{Abstract}
Serpentinization and carbonation of ultramafic formations is an ubiquitous phenomenon which deeply influences the biogeochemical cycles of water, hydrogen, carbon\ldots while supporting the particular biosphere around the oceanic hydrothermal vents. Carbonation of peridotites and other mafic and ultramafic rocks is also a hot topic in the current energy landscape as the engineered sequestration of mineral \ce{CO2} in these formations could help reduce the atmospheric emissions and cope with climate change. In this study, we present two reactive percolation experiments performed on a natural serpentinite dredged from the ultraslow South-West Indian Oceanic Ridge. The serpentinite cores (length 3-4 cm and dia. 5.6 mm) were subjected for about 10 days to the continuous injection of a \ce{NaHCO3}-saturated brine at respectively 160\textcelsius\ and 280\textcelsius. Petrographic and petrophysical results as well as outlet fluid compositions were compared to numerical batch simulations performed with the PHREEQC open software allowing to reconstruct the mineralogical evolution of both cores. The most striking observation is the fast and dramatic decrease of the permeability for both experiments principally due to the precipitation of carbonates. On the contrary, serpentine precipitation was found to be limited and less impacting as it precipitates in low-flow zones, out of the main percolation paths. In total, about 5.6\% of the total injected \ce{CO2} was retained in the core, at 280\textcelsius. In the same time, hydrogen was consistently produced with a total recovered \ce{H2} corresponding to 0.8\% of the maximum \ce{H2} possible considering the total amount of Fe(II) in the starting core.

The modeling of the results allowed to highlight the fundamental role of spatio-temporal lengthscales on the overall behavior of the core. These lengthscales are controlled by the competition between species transport and chemical kinetics scaled by \Da\ number. In particular, experimental results show that the behavior can drastically change depending on the pore size distribution since the variation of fluid velocity in the different classes of pores leads to different outcomes of the transport/kinetics competition.

\section{Introduction}
With global warming and increasing carbon dioxide concentrations in the atmosphere being probably one of the main scientific challenges of the XXI$^\text{st}$ century, the need to develop and implement Carbon Capture and Storage (CCS) technologies is more and more pressing. One solution advocated in the last 2 decades is the geological storage of \ce{CO2}, usually in large porous formations such as depleted gas and oil fields, deep saline aquifers or salt formations. Advantages of this technique are numerous, with in particular the immense storage capabilities of these repositories and the relative ease of injection \citep{Bachu2003}. However all these storage targets present the major shortcoming of potential leakage of the sequestrated \ce{CO2} to the surface \citep{Nelson2005} either by (1) leakage from the injection well due to issues with the cement casing and well completion \citep{Fabbri2012}, or (2) leakage from active or reactivated faults connecting the reservoir to the surface, or to groundwater \citep{Rohmer2010}. As \ce{CO2} is injected in supercritical form, and since its solubility in brine is very limited, any path opening toward the surface would lead to potentially dramatic consequences on the overlaying populations \citep{Guenan2011}. 

A potential solution is to store the injected \ce{CO2} under solid form so that no leakage is possible even in the presence of an open pathway to the surface. This solid form is achieved through the precipitation of carbonates (e.g. calcite $\ce{Ca^{2+} + CO2 + H2O = CaCO3 + 2 H^+}$). In the case of a deep saline aquifer or a depleted gas reservoir, despite the high salinity of the connate water, the quantity of available divalent cations necessary for this precipitation is too low to achieve a relevant yield of mineralization in a reasonable amount of time \citep{Johnson2004}. The solution advocated instead by several authors \citep{Seifritz1990,Kelemen2008} is to switch targets and inject \ce{CO2} into a chemically reactive formation where it would react with the host rock leading to mineralization in a much shorter time frame \citep{Gadikota2020}. Among the different options, mafic to ultramafic formations, rich in divalent cations (Mg, Fe, Ca) are probably the most attractive \citep{Oelkers2008} as basalts, peridotites and serpentinites are known to quickly react with carbon-rich fluids, precipitating carbonates, storing the \ce{CO2} safely over geological times \citep{Kelemen2018,McGrail2006}. Interestingly, these mineralization reactions are naturally occurring  in a variety of settings  -- e.g. as surface and near surface carbonation of ophiolites in Oman \citep{Kelemen2011}, Italy \citep{Boschi2009}, Norway \citep{Tominaga2017} or as higher temperature alteration as evidenced by the carbonate chimneys at hydrothermal vents -- leading to the formation for example of listvenite, a geological formation composed of a magnesite/quartz assemblage originating from the carbonation of ultramafic formation \citep{Hansen2005,Tominaga2017,Ulrich2014}.  Target formations are for example the ophiolitic complex in Oman which could, according to \citet{Kelemen2008}, store up to 77 trillion tons of \ce{CO2}, or basaltic formations in Iceland which have been the target for the first pilot scale injection of carbon dioxide with great success by the CarbFix project team \citep{Gislason2018}. 

Additionally, natural carbonation of mafic and ultramafic rocks is always tightly linked to serpentinization processes themselves \citep{Klein2013a,McCollom2020}, a specific set of reactions transforming the minerals of anhydrous mantle peridotites (i.e. olivine $\ce{Mg2SiO4}$, orthopyroxenes $\ce{Mg2Si2O6}$ and clinopyroxenes $\ce{MgCaSi2O6}$) into hydrated and less dense phyllosilicates  -- serpentine $\ce{Mg3Si2O5(OH)4}$, talc $\ce{Mg3Si4O10(OH)2}$ or clay minerals (sepiolite, saponite\ldots) -- through the influence of percolating seawater, usually in an oceanic settings, where ultramafic formations crop out next to slow spreading ridges \citep{Cannat2010}. One of the byproducts of these reactions is molecular hydrogen \ce{H2} which is generated by the oxidation of ferrous iron to ferric iron and the associated reduction of water \citep{Mccollom2009}. Engineered carbonation of mafic and ultramafic rocks could then also potentially be coupled with molecular hydrogen \ce{H2} production, an economically inescapable component in the energy transition, but whose production means are currently tightly linked to fossil fuel usages \citep{InternationalEnergyAgency2019}. The concomitant industrial production of \ce{H2} and carbon dioxide sequestration in ultramafic settings is a young field of investigation (e.g. \citep{Wang2019}) but whose results are not only important for the potential economical exploitation of these reactions, but also for the global understanding of (1) the deep carbon and water cycles (serpentinization represents an estimated production of 10$^4$ to 10$^6$ t/year worldwide of \ce{H2}, with the impressive examples of the seafloor smokers \citep{Charlou2010} while carbonation of silicate minerals could account for 0.1Gt$_{\ce{CO2}}$  per year \citep{Gaillardet1999}); (2) the rheological and geochemical properties of the mantle-dominated lithosphere \citep{Escartin1997,Paulick2006} and (3) the particular biosphere of these ultramafic environments \citep{Fruh-Green2004}.

However, the analysis and a fortiori the mastering of these geochemical reactions requires the understanding of the complicated coupling between the dissolution/precipitation chemistry and the transport of fluid and dissolved species through the porosity. Indeed, the injection/percolation of a carbonated brine in a geological formation triggers numerous chemical reactions, the occurrence and intensity of which depends on the local chemical equilibria and thus on the transport of heat and matter by the percolating fluid. In particular, the increase of molar volume from anhydrous to hydrated silicates (density of a peridotite is 3.3 g/cm$^3$ while it is 2.6 g/cm$^3$ for a serpentinite) leads to the clogging of the percolation path, clogging potentially counteracted by reaction-induced fracturing where the confined growth of minerals generates stresses on the confining porous matrix leading to damage, fracturing and even percolation paths opening  \citep{Jamtveit2000,Iyer2008,Kelemen2012,Malvoisin2017}. This dichotomy of clogging versus damage is fundamentally controlled by local, microscale processes and in particular by the location of the coupled dissolution/precipitation processes and their intricate relationship with advection and dissolved species transport \citep{Osselin2016,Osselin2019a}. A careful and extensive control on the reaction and injection characteristics as well as a precise control on the different chemical and physical processes is then required in order to understand these fundamental reactions and allow the development of this very promising technology. 

While batch experiments and reaction path studies are well described in the literature \citep{Klein2011,Grozeva2017,McCollom2020,Miller2017,Malvoisin2012,Neubeck2011}, percolation studies featuring solid cores, in which a reactive solution is being injected at constant flow rate or constant pressure drop, are more scarce. Most of these experiments \citep{Andreani2009,Peuble2015,Peuble2015a,Peuble2018,Peuble2019} for carbonation and \citep{Godard2013,Escario2018} for serpentinization were performed on sintered olivine in which carbonated water or artificial seawater respectively was injected at flow rates ranging from 0.1 to 12 ml/hr. Results highlighted a sharp drop in permeability (except for \citet{Andreani2009} and \citet{Escario2018} who on the contrary observed a constant permeability). This sharp drop is the consequence of the precipitation of neoformed minerals (respectively carbonates and serpentine or proto-serpentine) but also of the coupling between hydrodynamics and reaction kinetics, as slow precipitating minerals were evidenced to precipitate in low fluid velocity areas (secondary flow paths) while silicates tend to precipitate towards the outlet of the core. The usage of an artificial, ideal substrate (sintered olivine) and the somewhat limited temperature range of these experiments (160-185\textcelsius) unfortunately limits the applicability of the results. \citet{Luhmann2017} on the contrary used intact dunite cores for their two experiments at 150\textcelsius\ and 200\textcelsius\ with artificial seawater. They also highlighted a permeability drop attributed to the precipitation of silicates in the porous system. Additionally, this experimental set-up was also used to study the silicification of brucite \citep{Tutolo2018} at 150\textcelsius\ with a modified seawater injected at 0.05 ml/min. Finally, \citet{Farough2016} described the behavior of ultramafic cores, artificially fractured in two halves, and subjected to the percolation of deionized water at 260\textcelsius. They describe the loss of permeability during the experiment due to the precipitation of silicate minerals healing the central fracture. 

In this study, we present two original experiments on a natural partially-serpentinized peridotite subjected to the injection of a carbonated fluid at 160\textcelsius\ and 280\textcelsius, which corresponds respectively to a temperature close to the optimum for ultramafic mineral carbonation \citep{OConnor2005}; and close to the optimum of the serpentinization reaction \citep{Malvoisin2012,Martin1970,Marcaillou2011}. Compared to the current literature on the topic, these experiments present two original novelties. First, the starting material is a natural partially serpentinized peridotite containing olivine, clino- and orthopyroxenes as well as serpentine and aragonite, contrary to the ideal sintered olivine or even dunite used in the literature. This type of rock is representative of most deep ocean ultramafic formations (e.g. at ultraslow spreading ridges for example) and is thus better suited for in-situ carbonation and hydrogen production studies. Secondly, the second experiment was performed at 280\textcelsius, a common temperature of this kind of alterations \citep{Menzel2018} but which had not been investigated before. 

The goal of these experiments was to track the evolution of the petrophysical and geochemical characteristics of the cores and assess their potential for carbon dioxide sequestration and hydrogen generation, as well as analyze the complex reactive transport couplings of these reactions in natural settings.

\section{Materials and Methods}

\subsection{Starting materials and characterization}

\paragraph{Petrology} Three different thin sections were extracted from the initial serpentinite and analyzed with optical microscopy, SEM imaging coupled with EDS analyses, Electron Microprobe, Raman spectroscopy, as well as optical cathodoluminescence and micro-X-rays fluorescence ($\mu$XRF). Modal composition was obtained from point counting (0.5mm $\times$ 1 mm steps), including the 5 main identified phases: olivine, ortho- and clinopyroxenes, serpentine, aragonite and spinel. Magnetite was not included in the point counting due to the small size of the crystals. Similarly, aragonite was difficult quantify due to the thinness of the veinlet network. In order to obtain the msot accurate composition of the starting material, the composition from point counting was rescaled using the whole rock composition (SARM - Service d'Analyses des Roches et des Minéraux, Nancy -- see Supplementary information for details of the analytical procedures), as well as the EPMA compositions of the phases of interest. First aragonite was recalculated from the \ce{CO2} content and clinopyroxene were rescaled from the measured CaO minus aragonite. Serpentine, olivine and orthopyroxene contents were then tweaked to get the closest match for MgO and \ce{SiO2}, and finally, magnetite was obtained from \ce{Fe2O3}$_T$, taking into account FeO from the other phases. The final rescaled composition shown on Table \ref{tab:mineralogy_serp} remains very close to the point counting results and gives very accurate results when compared with the measured whole rock composition. In particular, spinel was kept identical to point counting as the \ce{Al2O3} content was already almost perfectly predicted. The only real difference is in the water content which is significantly overestimated. This is due to the calculation of the serpentine formula from EPMA assuming 1 \ce{H2O} per formula unit. The comparison with measured water in the sample suggests that the serpentine contains less water than the ideal formula. 

\paragraph{Petrophysical properties} In order to characterize the petrophysical characteristics of the starting rock, Mercury Intrusion Porosimetry (MIP) and BET analysis were performed on several cored samples. MIP was performed with an AutoPore IV 9500 apparatus from Micromeritics while BET analysis was performed on a Quantachrome Nova apparatus. Pre-experiment permeabilities were also measured for the two cores at room temperatures and at working pressures (i.e. 100 bar of outlet pressure for the 160\textcelsius\ experiment and 200 bar for the 280\textcelsius\ experiment -- confinement pressure was set at least 70 bars above the injection pressure). The measurement consisted in setting a pressure difference ($\Delta P$) between the inlet and outlet and measuring the steady-state flow rate. In order to improve the reliability of the measurements, a total of 4 ($\Delta P$,Q) pairs were performed for each core and the permeability was retrieved from the slope of the linear regression \citep{Osselin2015} using Darcy's law -- $Q/A = (\kappa/\eta) (\Delta P/l)$ with $Q$ the measured flow rate, $A$ the cross section of the core, $\eta$ the viscosity of the fluid at the considered pressure and temperature, $\Delta P$ the imposed pressure difference and $l$ the length of the core. Since these pre-measurements were performed at room temperature, no chemical reaction was expected to occur. Results yielded a value of approximatively 2mD for the core used for the experiment at 160\textcelsius\ and 4mD for the core used for the experiment at 280\textcelsius. These values are rather high, especially for this kind of rock \citep{Reynard2013,Hatakeyama2017} and are due to the large proportion of open fractures in the cores (1mD = 1.10$^{-15}$m$^2$).

\subsection{Reactive percolation experiments}
The experiments started from cores ($\varnothing$ 5.6 mm, lengths 4.1 cm for the experiments at 280\textcelsius\ and 3.9 cm for the experiment at 160\textcelsius) drilled from a serpentinite block dredged from the South-West Indian Ocean ridge \citep{Roumejon2014}. The injected solution was prepared in advance as 2wt\% NaCl and 5wt\% \ce{NaHCO3} (i.e. $\approx$ 0.34M \ce{NaCl} + 0.6M \ce{NaHCO3}) with Sodium Chloride from Alfa Aesar (99+\% purity) and Sodium bicarbonate from SigmaUltra (99.5\% purity). Room temperature pH was around 8. This composition is close to the potential conditions relevant to \ce{CO2} storage \citep{Gadikota2020} as well as average natural water hosted by ultramafic formations (albeit at lower temperatures) \citep{Cipolli2004}. Moreover, as highlighted by \citet{Lafay2018}, \ce{NaHCO3}-rich fluids used for experiments (e.g. \citet{OConnor2005,Hovelmann2011,Peuble2015a}), have proven to effectively reproduce  alteration features from natural serpentinization/carbonation reactions, while allowing for reasonably rapid reaction rates.

\begin{figure}
	\centering 
	\includegraphics[width=0.89\textwidth]{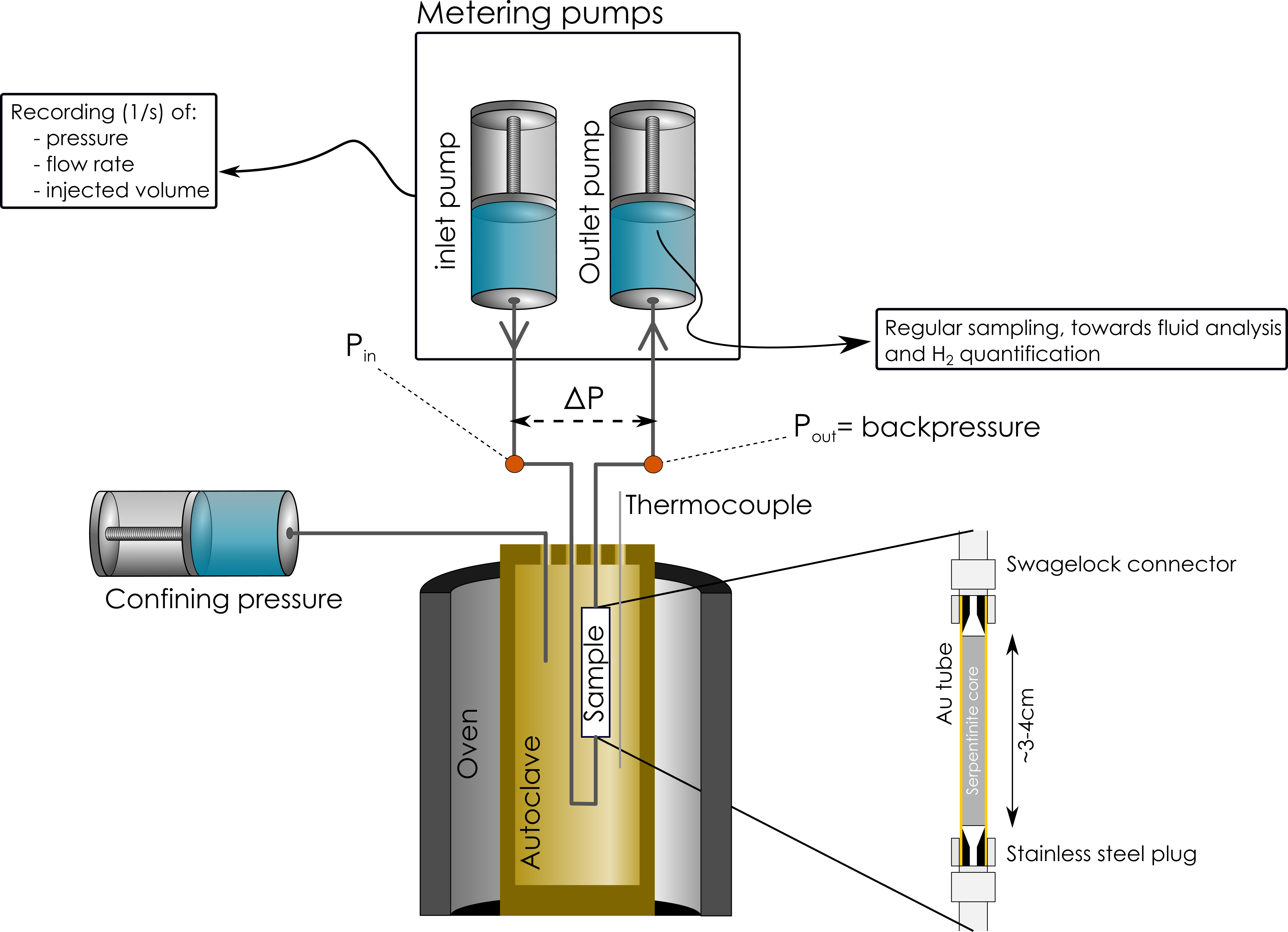}
	\caption{\label{fig:montage_total} Schematics of the permeameter}
\end{figure}

The two reactive percolation experiments (one at 160\textcelsius\ denoted as LTE - Low Temperature Experiment and the other at 280\textcelsius\ denoted as HTE - High Temperature Experiment) used a home-made permeameter represented on Figure \ref{fig:montage_total}. It consists in a large volume (1 liter) autoclave connected to a piston-pump (PMHP 100-500, TOP Industry) controlling the confining pressure using deionized water as the confining fluid. The core itself is inserted in a gold tube working as a jacket which transmits the confining pressure. At both ends of the core, two stainless steel plugs are added to connect with the inlet and outlet capillaries using Swagelok tube fittings as represented in the inset of Figure \ref{fig:montage_total}. The conical shape of the plugs allows for an even distribution of the flow at the inlet and outlet. The sample is connected to two more pumps  controlling the inlet and outlet pressure. The autoclave is heated externally by a cylindrical oven controlled with an Eurotherm regulator. Experimental temperature is measured inside the vessel next to the core using an internal sheathed type K thermocouple and is known with an uncertainty < 10\textcelsius. Temperatures fluctuations during the experiments were limited to 1-2\textcelsius.

The system works at temperatures up to 400\textcelsius\ and confining pressures up to 500 bar. In this study, the outlet pressure was set at 100 bar for LTE and 200 bar for HTE. Confining pressure was set 70 bar above the inlet pressure. As clogging during the experiments occurs fast, we decided to control the injection by imposing a constant pressure difference between the inlet and the outlet of the core in order to avoid an uncontrolled increase of pressure in the case of complete clogging. The pressure difference was initially set at 2 bars and was gradually increased during the experiment following the decrease of permeability to maintain a flow rate around 0.5-0.6 ml/hr. Resulting flow rate as well as inlet/outlet pressures and temperature were recorded with a frequency of 1 data point per second for the whole duration of the experiment using an interfacing with the software Eurotherms iTools. The time-dependent permeability is then calculated from Darcy's law as the ratio between the imposed pressure difference between inlet and outlet and the corresponding measured flow rate. Viscosities were calculated from the IAPWS-97 equation as 2.10$^{-4}$ Pa.s at 160\textcelsius\ and 1.10$^{-5}$ Pa.s at 280\textcelsius

	\paragraph{Fluid analysis}
During HTE, for time intervals of roughly 24 hours at the beginning and less frequently thereafter, the outlet pump was emptied and the outlet fluid was sampled for chemical analysis. Samples were immediately filtered on a 0.45$\mu$m nylon filter, stored in a fridge at 4\textcelsius\ and quickly analyzed, without any acidification. No water analysis was performed for LTE. Cation analysis was performed at CETRAHE (Ionic chromatography on a Dionex ICS 1100) while anion analysis was performed on site on a Ionic Chromatography with a Dionex Ionpac AS11-HS column, and  DIC (Dissolved Inorganic Carbon) was analyzed with a Shimadzu TOC-L series.  Dissolved hydrogen was also analyzed by gaz chromatography (see Supplementary materials 1 for the sampling and analytical protocol). 

	\paragraph{Mineral analysis}

After the end the experiment, the core was retrieved and cut axially in half. A polished section of each half-core was prepared for SEM examination coupled with EDS analyses, cathodoluminescence, Raman spectroscopy, micro-X-rays fluorescence and microprobe analyses.

\subsection{Numerical simulations with PHREEQC}
In order to analyze the chemical evolution during the experiments, numerical simulations of equivalent batch experiments  were performed with PHREEQC \citep{Appelo2015} and the Thermoddem database \citep{Blanc2012} under LTE and HTE pressure and temperature conditions (resp. 160\textcelsius, 100 bar and 280\textcelsius, 200 bar). The simulations were run as 1 kg of starting material to which was added at each step 1 pore volume of the inlet solution (2\% NaCl, 5\% \ce{NaHCO3} as in the actual experiments) considering a porosity of 10\%. After equilibration, the solution is replaced by another pore volume of fresh solution until a total of 5000 pore volumes has been injected. This procedure simulates the continuous injection of the solution in the core considering infinitely fast chemical reactions. 

The starting material composition used for the simulations is the rescaled composition from Table \ref{tab:mineralogy_serp} without the spinel phase. Indeed this phase appears mostly undisturbed after the experiments and its addition in the simulations would result in unnecessary complexity. All the starting minerals are considered as pure phases of fixed composition as determined by EPMA (Supplementary information 3) and are only allowed to dissolve. Their thermodynamic properties were calculated using the thermodynamic properties of their end-member components and assuming ideal mixing behavior. The Thermoddem database does not incorporate a description of the influence of Al on the thermodynamic properties of pyroxenes and, so, the presence of Al in the proxenes was not taken into account. 

For the precipitation, two kinds of product phases were considered, (1) solid solutions (assumed to behave ideally), such as neoformed serpentine, carbonates\ldots\ and (2) pure phases such as quartz or hematite (see Supplementary Information 3 for the complete list). In order to describe Fe(III) incorporation in the serpentine solid solution, we used the data for cronstedtite-7\AA\ (\ce{Fe2Fe2SiO5(OH)4}) from the CarbFix database \citep{Voigt2018} instead of the Thermoddem database where Al is present. The Al substitution in serpentine is instead modeled using data for amesite, \ce{Mg2Al(AlSiO5)(OH)4}, the Al-bearing serpentine end-member. This phase, which is structurally closer to lizardite, is preferred to kaolinite, which was used in some previous modelling \citep{Klein2013,Malvoisin2015}. 

A few other phases such as some amphiboles (edenite, riebeckite\ldots) and sodic pyroxenes (aegirine) were supersaturated in some of the simulations. However, these phases were not detected in the experiments and are not included as potential product phases. In preliminary simulations, hydrated magnesium carbonates (hydromagnesite and nesquehonite since there is no data for dypingite) were considered as stable phases. However, nesquehonite, regularly the predominant phase at high temperatures, is supposed to  dehydrate beyond 100-150\textcelsius\ \citep{Morgan2015}. Therefore, both hydromagnesite (which is never stable in the simulations) and nesquehonite were subsequently removed from the potential list of stable solids. Saponite and vermiculite are the only clay minerals considered in the simulations. For simplicity, pure Na end-members are considered in the simulations despite their large compositional variability, and also because of the large Na activity in the initial solution. For Fe-oxides, magnetite and hematite and, for silica, quartz were considered as potentially saturating phases. For the carbon species, reduction of \ce{CO2} to \ce{CH4} by \ce{H2} was assumed not to occur. Indeed, this reaction is not observed in most experiments from the literature, being considered as kinetically inhibited \citep{Mccollom2001}. However, reduction of carbon dioxide into organic acids or even in some cases to solid carbon has been sometimes reported  \citep{Andreani2019,Peuble2019}. Nevertheless, given that definitive information is lacking on reaction paths and thermodynamical constants, such reactions were not considered, even if the saturation index of graphite was tracked in the different simulations.

\section{Results}

	\subsection{Initial serpentinite}
	
	\subsubsection{Petrography}
\begin{figure}
	\centering 
	\includegraphics[width=0.80\textwidth]{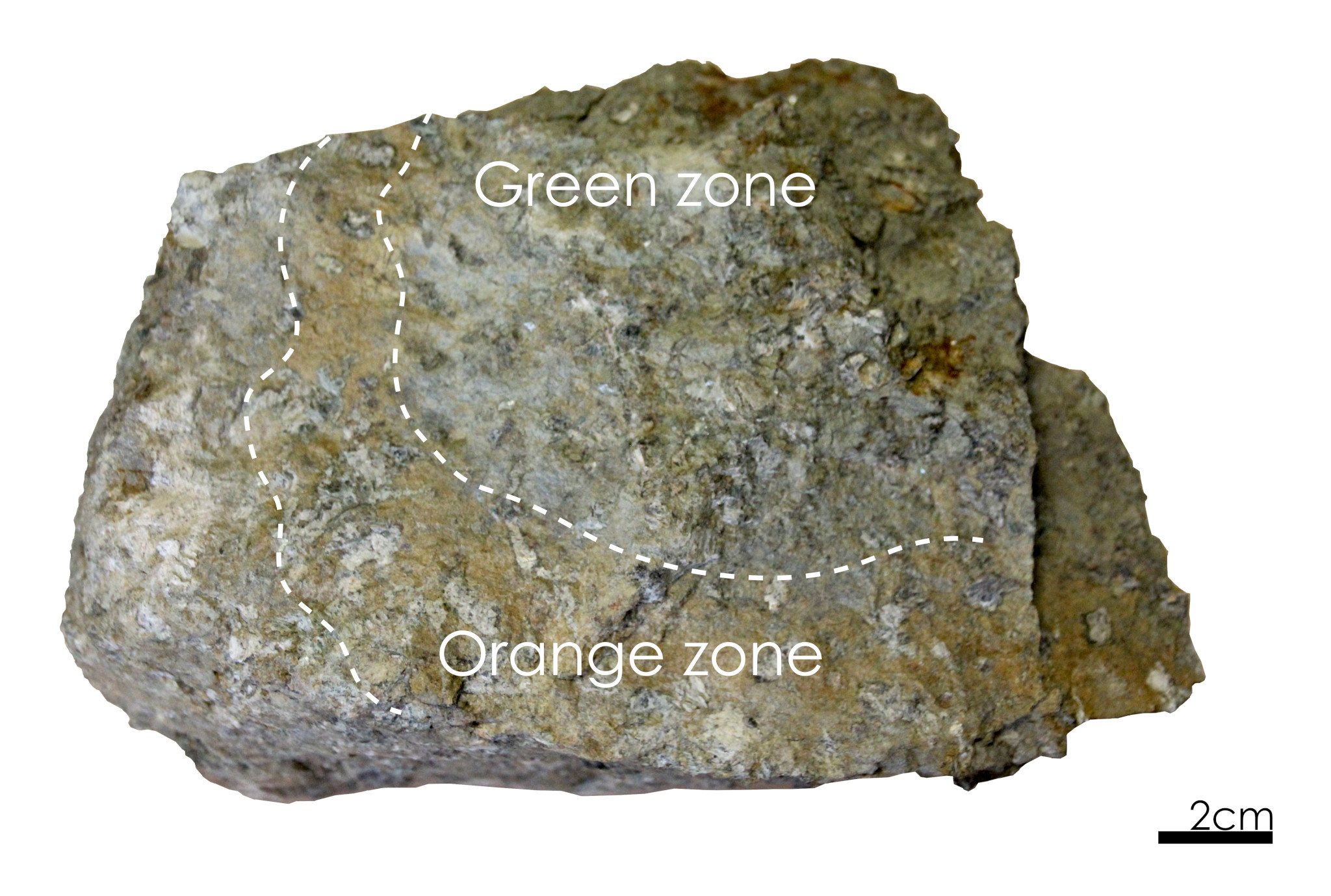}
	\caption{\label{fig:greenorange}Macroscopic view of the initial serpentinite. The green zone is characterized serpentinization features with little to no carbonates, while the orange brown zone is characterized by pervasive carbonation}
\end{figure}

\begin{figure}
	\centering \includegraphics[width = \textwidth]{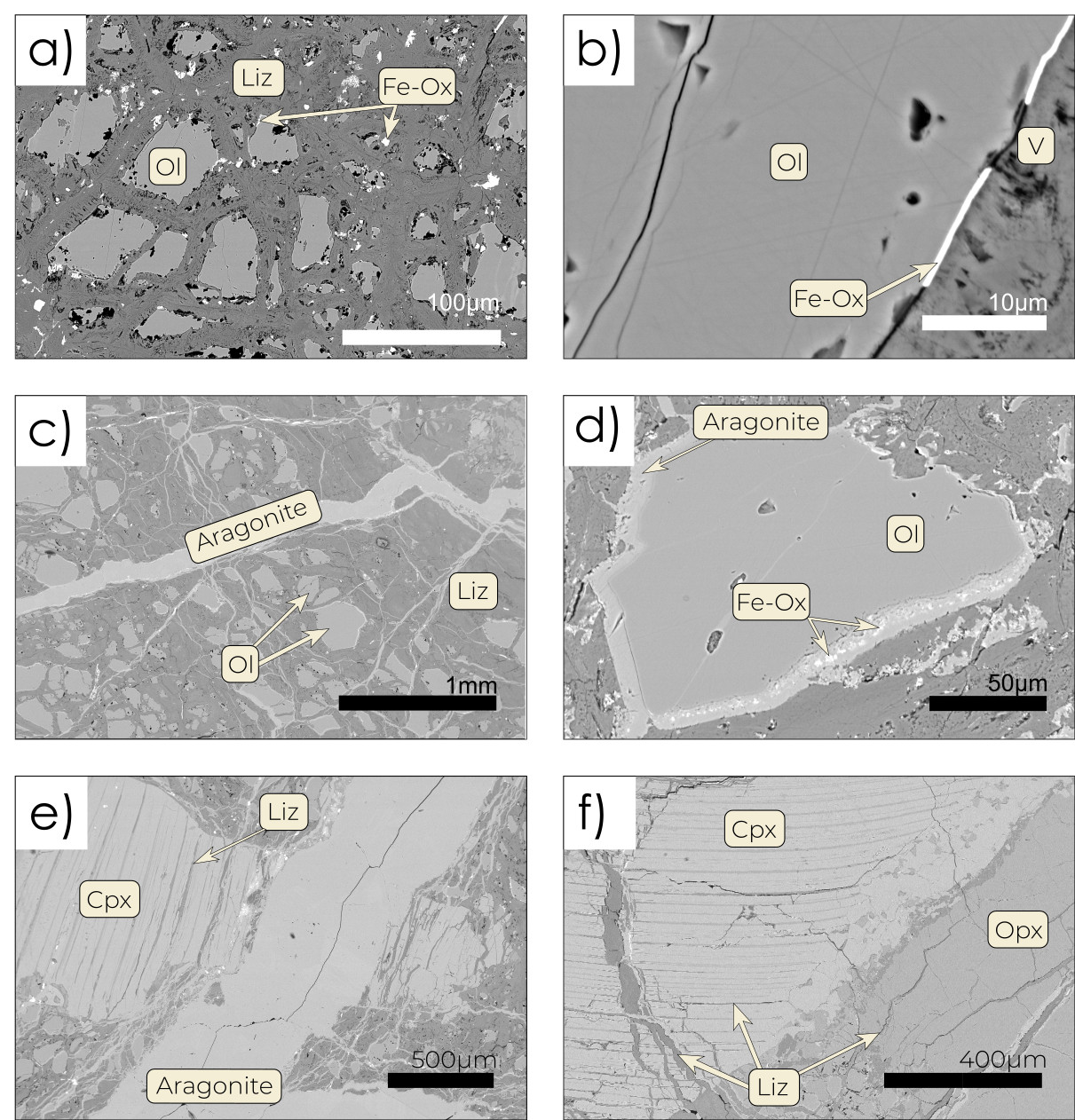}
	\caption{\label{fig:serp_init} SEM images of characteristic textural features in the starting serpentinite. a) View of the green zone showing the olivine relicts and the serpentine matrix with large Fe-oxides. b) enlarged view of a) showing the olivine-matrix interface with small locally distributed Fe-oxides. Notice the void at the interface. c) Image of the  brown orange zone showing different generations of carbonate-filled veins crosscutting the matrix. d) enlarged view of c) showing the continuous carbonate layer with Fe-oxide inclusions around the olivine. e) large aragonite vein ($\approx$ 1 mm) cutting through an altered clinopyroxene. The alteration follows the cleavages and is made of serpentine (bastite). f) clinopyroxene (left with alteration along the cleavage and some fractures, and orthopyroxene (right) showing comparatively little alteration and only along fractures. Notice the contrast in alteration textures between, in the green zone (a, b), the thin and discontinuous iron oxide rim around the olivine, and, in the brown orange zone (c, d) the continuous layer of carbonate. \\
	 \emph{Cpx=clinopyroxene, Ol=Olivine, Serp=serpentine, Fe-Ox=Iron oxides and V=void}}
\end{figure}

The starting material is a partially serpentinized lherzolite containing olivine, ortho- and clinopyroxenes and serpentine as major minerals plus several secondary phases (aragonite, magnetite and spinel). The sample shows classic features of serpentinization on which carbonation events are superimposed (carbonates are identified as aragonite on XRD diffractograms and Raman spectra). It comprises two parts, both rich in olivine, which alternate at the cm scale and are best distinguished by their contrasted colors, respectively green and brown to dark orange (Fig. \ref{fig:greenorange}). The green zones contain carbonates only as large veins (mm to cm-scale), and are made up of relictual olivine in a serpentine mesh (Fig \ref{fig:serp_init}a). Oxides (magnetite) are abundant, generally of small size and they are typically found partially rimming olivine crystals in the void at the interface with the surrounding serpentine matrix (Fig \ref{fig:serp_init}b). Other occurrences of oxides are found as small nodules embedded in the serpentine matrix ($\approx$ 10$\mu$m) or in larger veins, a few $\mu$m wide and about 100$\mu$m long, visibly filling already existing cracks in the serpentine matrix. Green areas are also visibly porous as microfractures and sometimes mm-sized fractures develop in the serpentine matrix. 

\begin{figure}
	\centering \includegraphics[width=\textwidth]{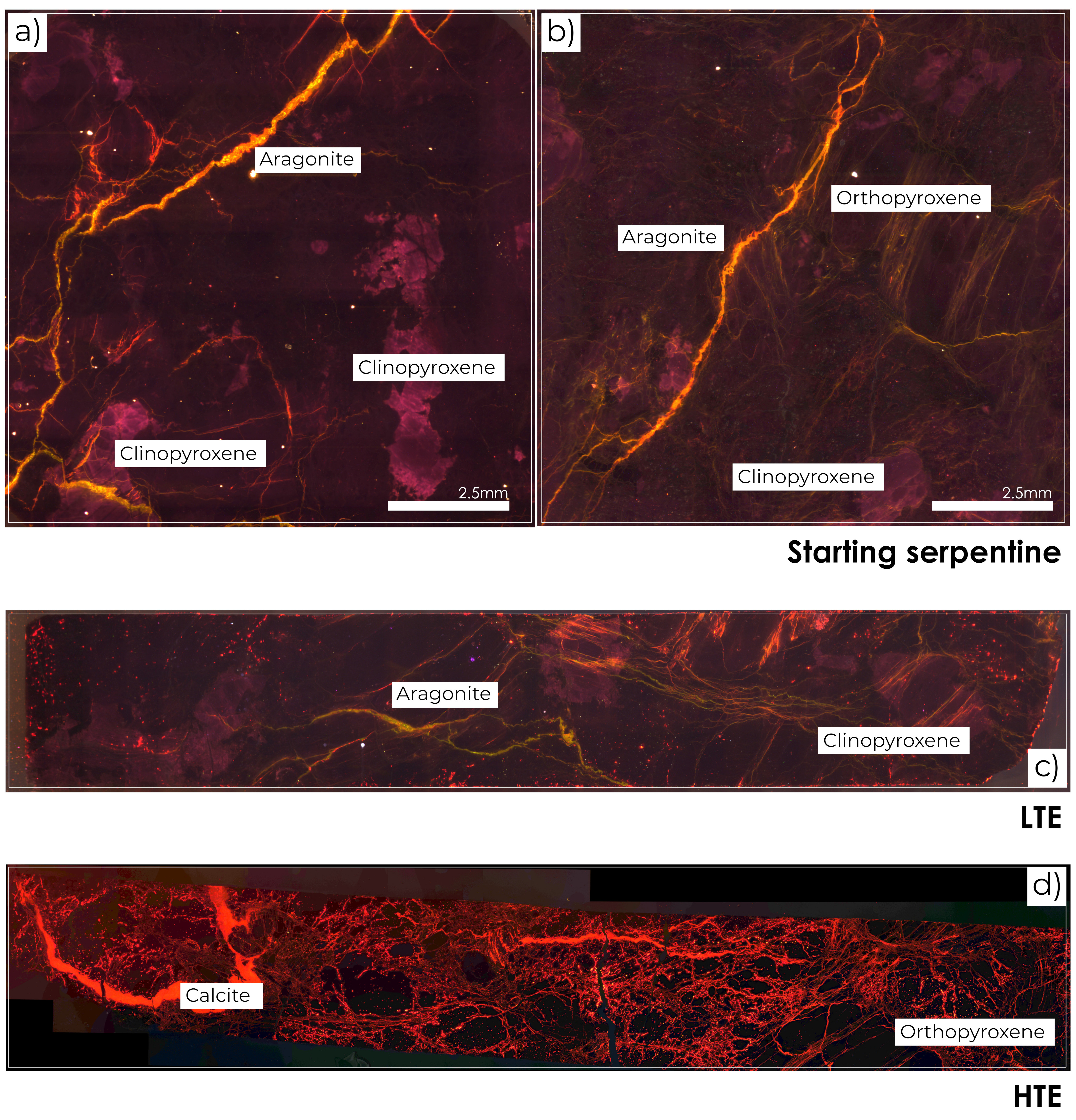}
	\caption{\label{fig:cathodo_image} Cathodoluminescence images from the starting material (a and b), LTE (c) and HTE (d). Note the stark contrast between a,b,c versus d. In the starting material as well as LTE, aragonite is visible as a loose network, appearing yellow to green and presenting little luminescence. In the contrary, calcite on HTE is very luminescent with a very dense network of red and bright luminescence.}
\end{figure}

In contrast, the brown to dark orange zones are associated with more pervasive carbonates and a much smaller quantity of Fe oxides (Fig. \ref{fig:serp_init}c,d). The latter occur embedded in the serpentine matrix ($\approx$ 1$\mu$m) or in carbonate-rich areas. Carbonates in the brown to dark orange zone appear to have cemented most of the porosity visible in the green zone. A complete cementation of the space between olivine and matrix is observed (Fig. \ref{fig:serp_init}d) as well as the occurrence of numerous thin veinlets connecting olivine relicts, probably originating from fracturating events with olivine acting as hard spots around which stresses accumulated. Carbonation of the protolith postdates serpentinization, with veins of carbonate (aragonite) crossing and overprinting veins and serpentine structures. In particular, the large vein in Figure \ref{fig:serp_init}e, is the result of the filling of a large fracture which appeared after most of the serpentinization occurred since alteration features can be matched from both sides of the vein. The center of this large vein appears white under the microscope and is analyzed as pure \ce{CaCO3} aragonite with EDS and Raman spectroscopy, while the borders present a rusty color visible with naked eye and under the microscope. These borders appear to consist in an intimate mixture of Mg and Ca-carbonates and small oxides. Aragonite under cathodoluminescence offers a color palette from yellow to green through light orange, but with overall little luminescence (Fig. \ref{fig:cathodo_image}a,b). The silicate background is significantly less luminescent with a dark pink shade, where only cpx can be discerned due to their lighter pink color. 

In both zones, olivines present a clear hierarchical pattern  -- a pattern resulting from the successive fracturing of the initial mineral into smaller domains \citep{Iyer2008} -- due to the fracturing of the olivine core and the replacement along the fractures (e.g. \citet{Malvoisin2017}). Dissolution features are common but not ubiquitous and some olivine/serpentine interfaces appear quite smooth while etch pits and saw-toothed interfaces can be observed on other occasions. Pyroxenes show different levels of replacement. As a general rule, clinopyroxenes appear less altered than orthopyroxenes. The least afflicted clinopyroxenes present light alterations along the cleavage (with serpentine as the replacing phase) and which progress orthogonally to it as observed on Figures \ref{fig:serp_init}e and f. At more advanced levels of replacement, the symmetry with respect to the cleavage planes is lost and clinopyroxenes start breaking into smaller independent fragments. Replacement can also occur along fractures not necessarily following the cleavage planes (Fig. \ref{fig:serp_init}f). As a result, some clinopyroxenes seem to be replaced without any visible pattern while some more intact crystals present a lamellar structure as in Figure \ref{fig:serp_init}e. Orthopyroxenes present a different pattern of alteration with little to no replacement along the cleavage and an alteration progressing mostly from fractures running through the mineral (Fig. \ref{fig:serp_init}f). Finally, the thin sections reveal some occurrences of Cr-spinel with different degrees of alteration, and some very rare occurrences of iron sulfide (probably pyrite) which occur as large (20$\mu$m to 100$\mu$m diameter) blobs. 

From the EPMA results, two types of serpentine can be distinguished in both zones. An Al-poor serpentine (Mg\#$\approx$90, \ce{Al2O3}$\approx$0.2wt\%), is identified as lizardite by Raman spectroscopy and makes up the mesh texture surrounding the olivine relicts and originates from olivine serpentinization. The other type of serpentine is also identified as lizardite but contains a higher Al content (\ce{Al2O3} 5-6wt\%, Mg\#$\approx$88) substituting for Si and is linked to the replacement of pyroxenes (bastite). On some rare occasions, veins of chrysotile can be observed which seem to have formed between the main serpentinization event and the carbonation event. The rest of the ultramafic minerals are forsteritic olivine Fo$_{90}$, enstatitic orthopyroxene (En$_{89}$) and clinopyroxene Di$_{91}$ (Di = X$_{Diopside}$ / (X$_{Diopside}$ + X$_{Hedengergite}$) ).

\begin{table}
\centering 
\begin{footnotesize}
\begin{tabular}{lccccccc}
\toprule
Mineral & Serpentine & Orthopyroxene  & Clinopyroxene & Olivine & Aragonite & Spinel & Magnetite \\
\midrule
Green Zone 1 (\%wt) & 49.6 & 15.6 & 10.0 & 9.2 & 13.4 & 2.1 &  -\\
\midrule 
Orange Zone 1 (\%wt) & 63.7 & 18.5 & 6.7 & 8.0 & 2.7 & 0.4 &  -\\
\midrule 
Orange Zone 2 (\%wt) & 52.9 & 20.8 & 16.7 & 7.0 & 0.9 & 1.7 &  - \\
\midrule 
Average (\%wt) & 50.2  & 19.7 & 12.9 & 9.4 & 5.9 & 1.9 &  - \\
\midrule
Rescaled composition (\%wt) & 51 & 20 & 11.5 & 10 & 4.3 & 1.7 - 1.5 \\
\bottomrule
\end{tabular}
\end{footnotesize}

\begin{scriptsize}
\begin{tabular}{lcccccccccccccc}
  & & & & & & & & & & & & & & \\
 \cmidrule[1pt]{1-3}
	\multicolumn{3}{c}{Whole Rock composition} & & & & & & & & & & & & \\
	\midrule
	Oxide(\%wt) & \ce{SiO2} & \ce{Al2O3} & \ce{Fe2O3}$_T$ & \ce{FeO} & \ce{MnO} & \ce{MgO} & \ce{CaO} & \ce{Na2O} & \ce{K2O} & \ce{TiO2} & \ce{P2O5} & LOI & \ce{CO2} & \ce{H2O}\\
	\midrule
	Measured & 41.41 & 3.07 & 8.29 & 4.74 & 0.11 & 33.31 & 5.03 & 0.15 & < ld & 0.091 & < ld & 7.69 & 1.88 & 6.38 \\
	\midrule
	Recalculated & 40.0 & 3.06 & 8.35& - &  - &  32.4 & 5.09 & - & - & - & - & - & 1.89 & 8.97 \\
	\midrule
	Relative error (\%) & 3.3 & 0.25 & 0.69 & - & - & - 2.6 & 1.2 & - & - & - & - & - &0.57 & 40.6 \\
	\bottomrule
	
	\end{tabular}
\end{scriptsize}
\caption{\label{tab:mineralogy_serp} Top table: Mass fractions of the different minerals calculated from point counting. Bottom table: measured and recalculated whole rock composition from the modal composition and the composition of the different phases as measured by EPMA. LOI = loss on ignition. Major elements quantified by IPC-OES. \ce{Fe2O3}$_T$ = FeO + \ce{Fe2O3}.}

\end{table}

\subsubsection{Petrophysical properties}

\begin{figure}
		\centering \includegraphics[width=0.45\textwidth]{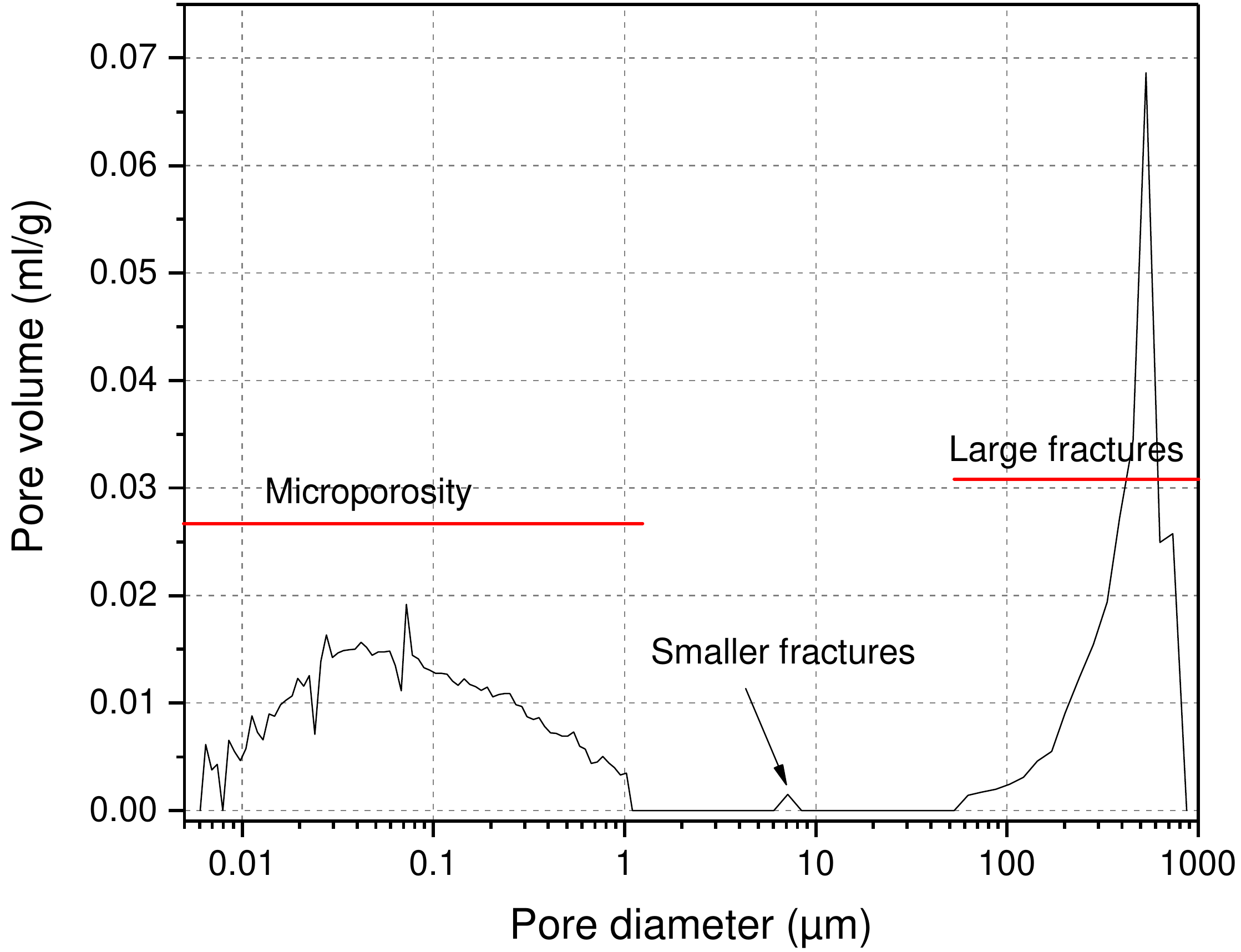}
		\caption{\label{fig:MIP} Pore size distribution in the starting serpentinite}
	\end{figure}

The mercury intrusion porosimetry distribution (Fig. \ref{fig:MIP}) is centered around two classes of pores, with one class consisting of submicronic porosity (visible on SEM images both as the space between olivine and serpentine as well as the microfractures in the serpentine matrix) and the other consisting in larger pores with sizes from around 60$\mu$m and higher, most likely larger fractures. A smaller third class of pores with sizes between 5-10$\mu$m is also visible but it represents only a small fraction of the pore volume. This class of pores is however rather important because it regroups a dense and connected network of small fractures, some of them partially filled by carbonates forming the so-called veinlets. In average, porosity of the cores is around 10\%. BET measurements provide a specific surface area of around 10-12 m$^2$/g which is divided into $\approx$ 7 m$^2$/g for the micropores (< 2 nm) and 3-5 m$^2$/g for bigger pores. Mercury intrusion porosimetry can as well provide an estimate of the specific surface area with 1.9.10$^{-4}$ m$^2$/g for pores between 871 $\mu$m and 62 $\mu$m, 3.3.10$^{-3}$ m$^2$/g for pores between 62 $\mu$m and 1 $\mu$m, and finally 2.3 m$^2$/g for pores below 1 $\mu$m. Since MIP does not consider pores below 6 nm, the rest of the specific surface area belongs to pores smaller than 6 nm which is consistent with BET results. 

\subsection{Numerical simulations} 
\begin{figure}
	\subfloat[ 160\textcelsius]{\label{fig:WR-160} \includegraphics[width=0.45\textwidth]{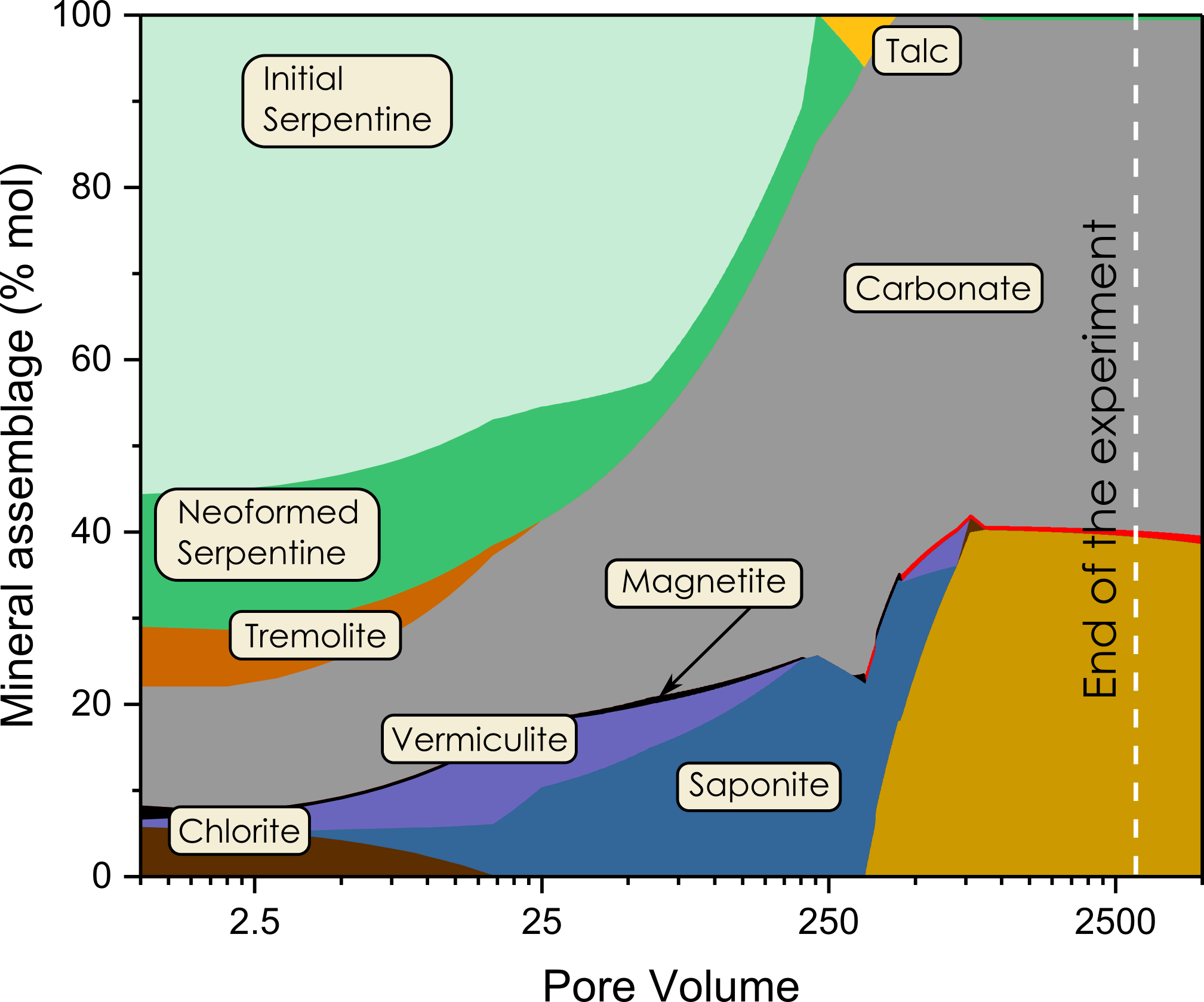}} \qquad	
	\subfloat[280\textcelsius]{\label{fig:WR-280} \includegraphics[width=0.45\textwidth]{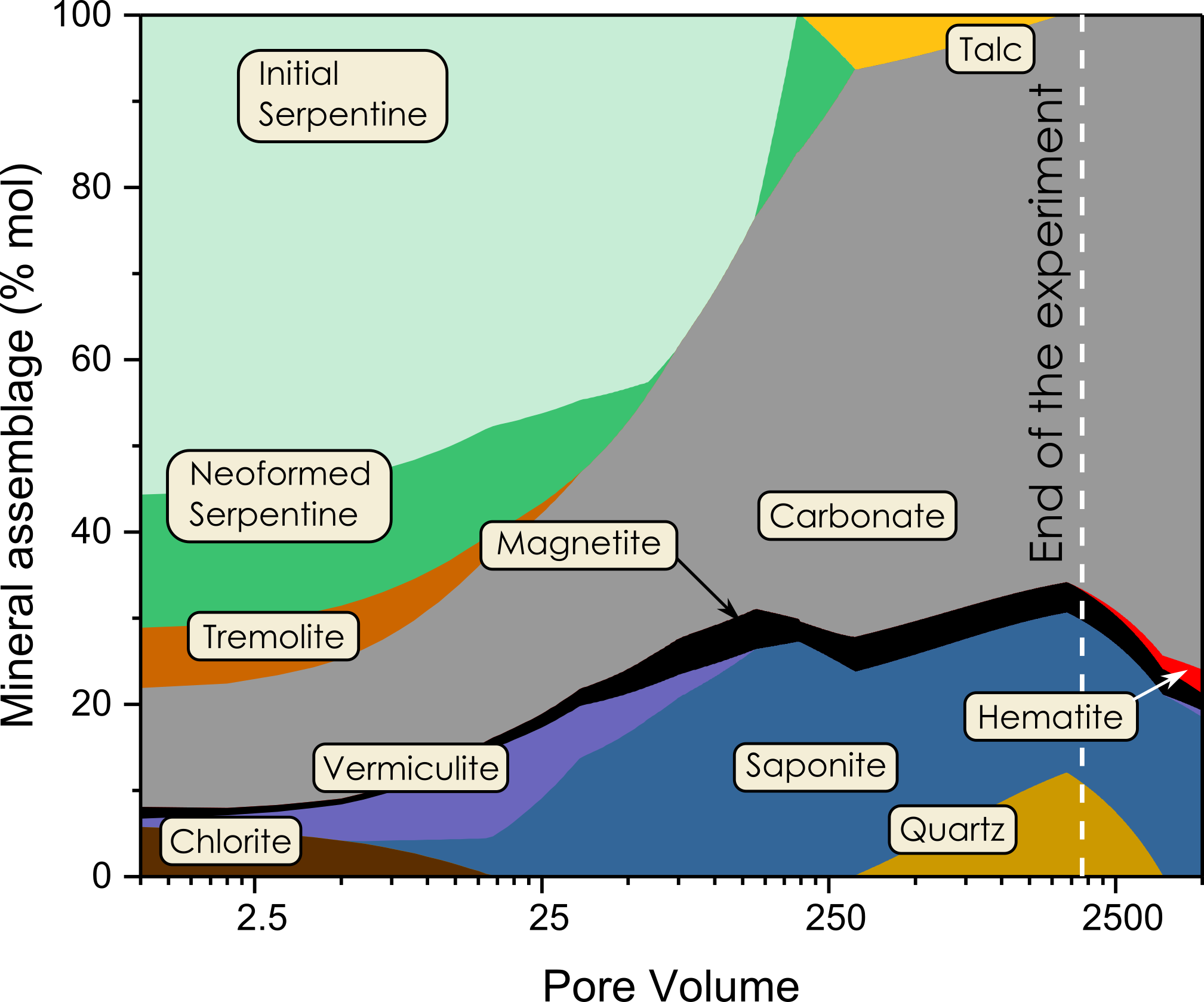}}	 \\
	\subfloat[160\textcelsius]{\label{fig:assemblage_serp_160} \includegraphics[width=0.45\textwidth]{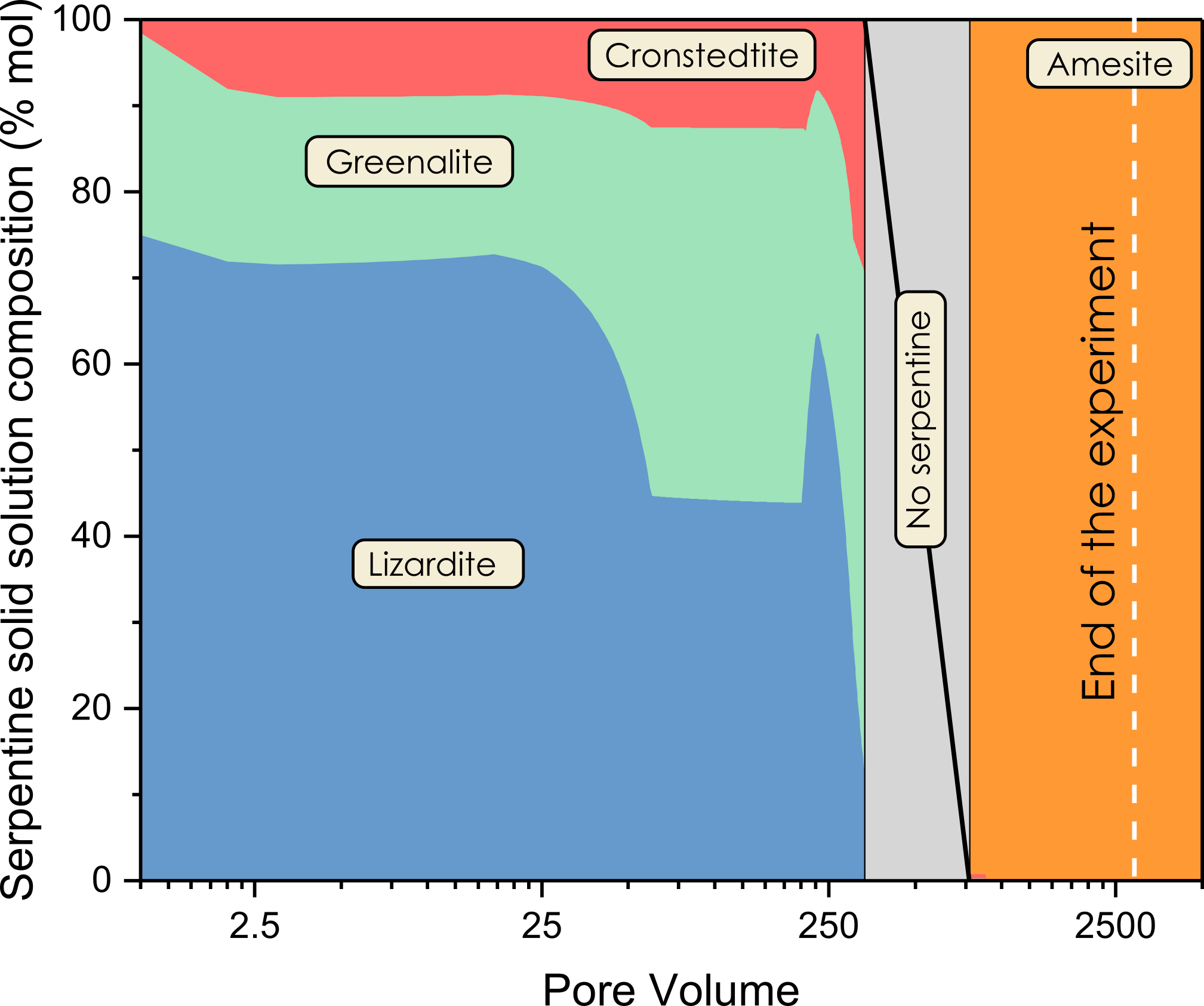}} \qquad	
	\subfloat[280\textcelsius]{\label{fig:assemblage_serp_280} \includegraphics[width=0.45\textwidth]{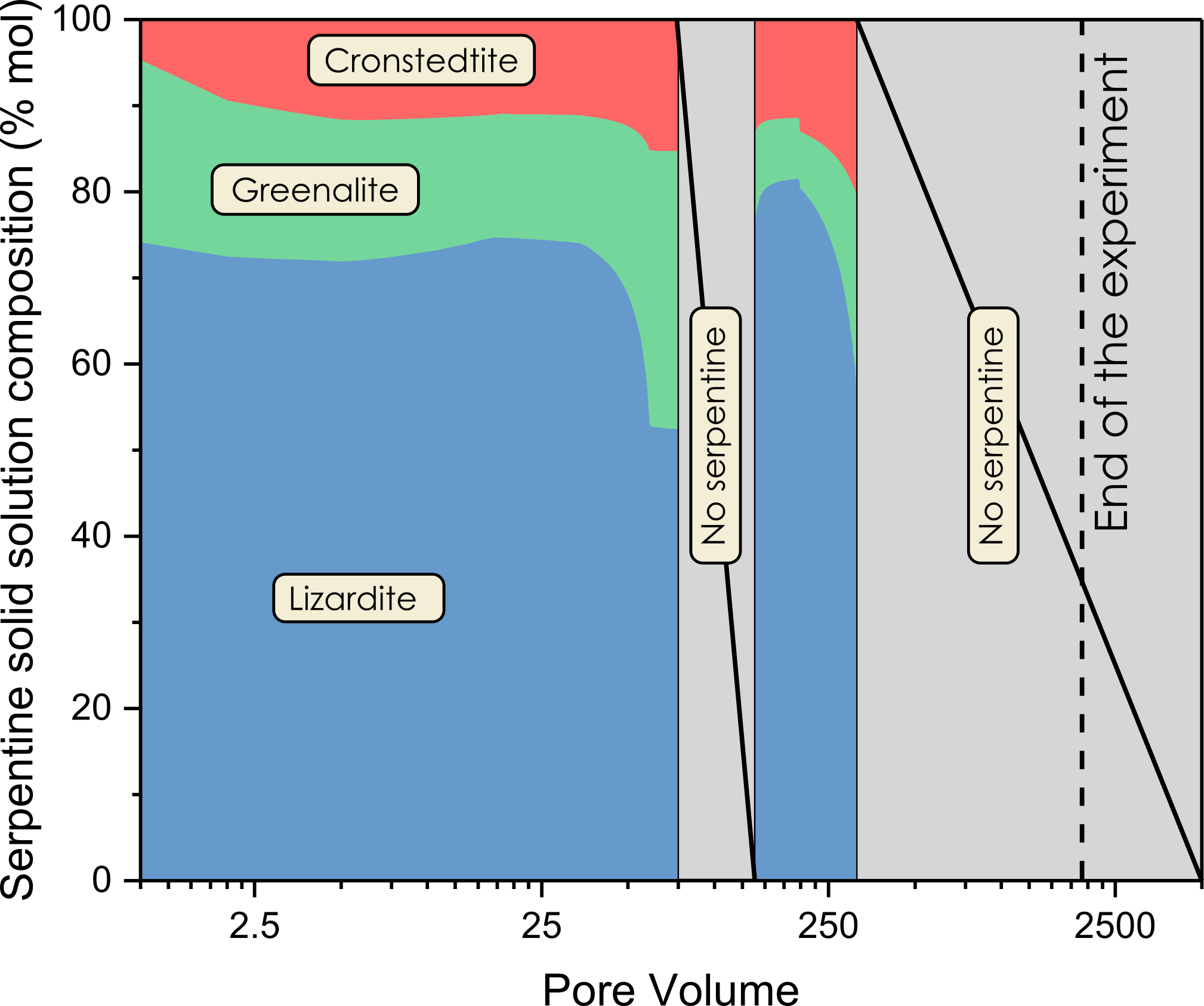}}	 \\
	\subfloat[160\textcelsius]{\label{fig:assemblage_increment_carbo-160} \includegraphics[width=0.45\textwidth]{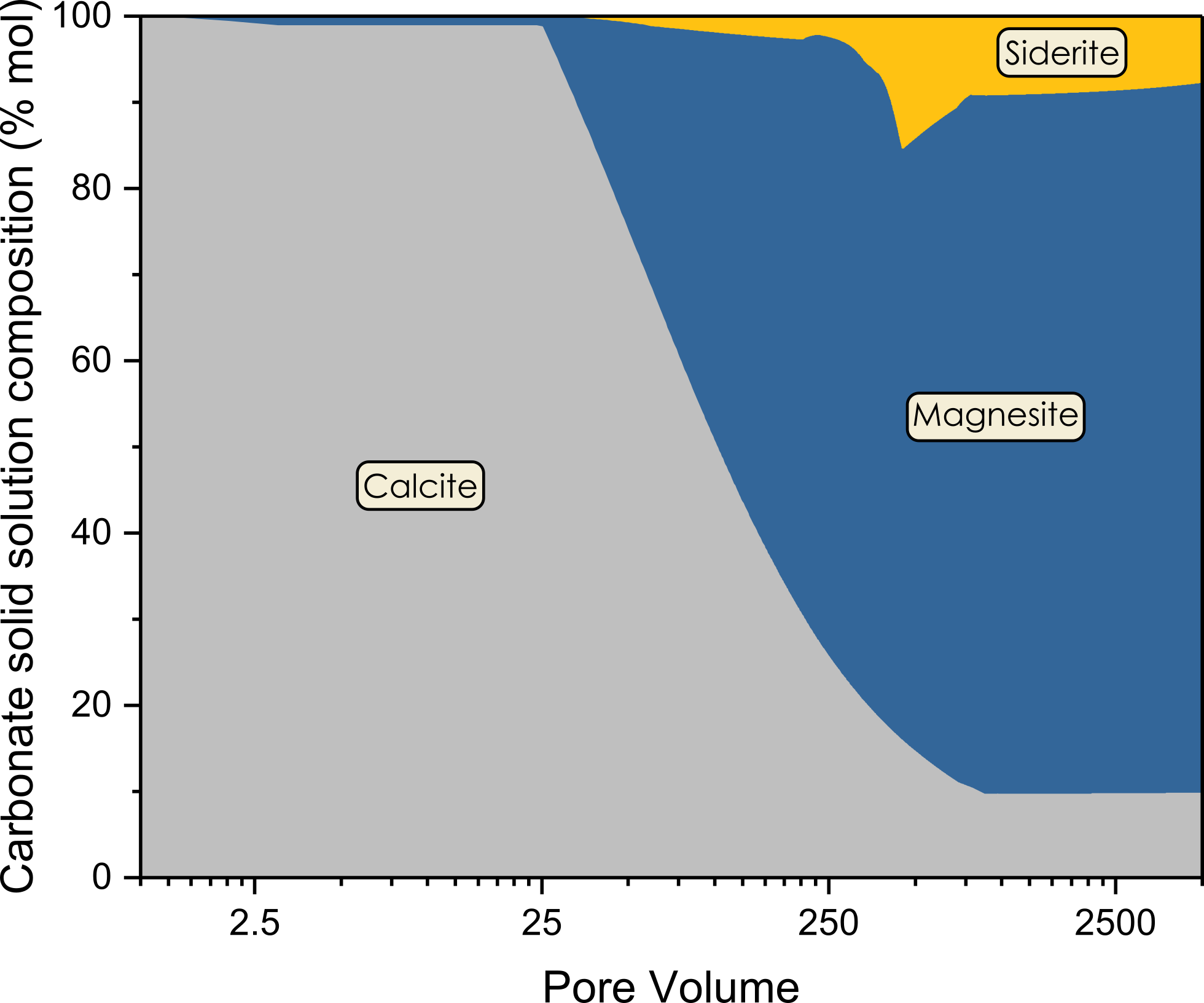}}
		\qquad
	\subfloat[280\textcelsius]{\label{fig:assemblage_increment_carbo-280} \includegraphics[width=0.45\textwidth]{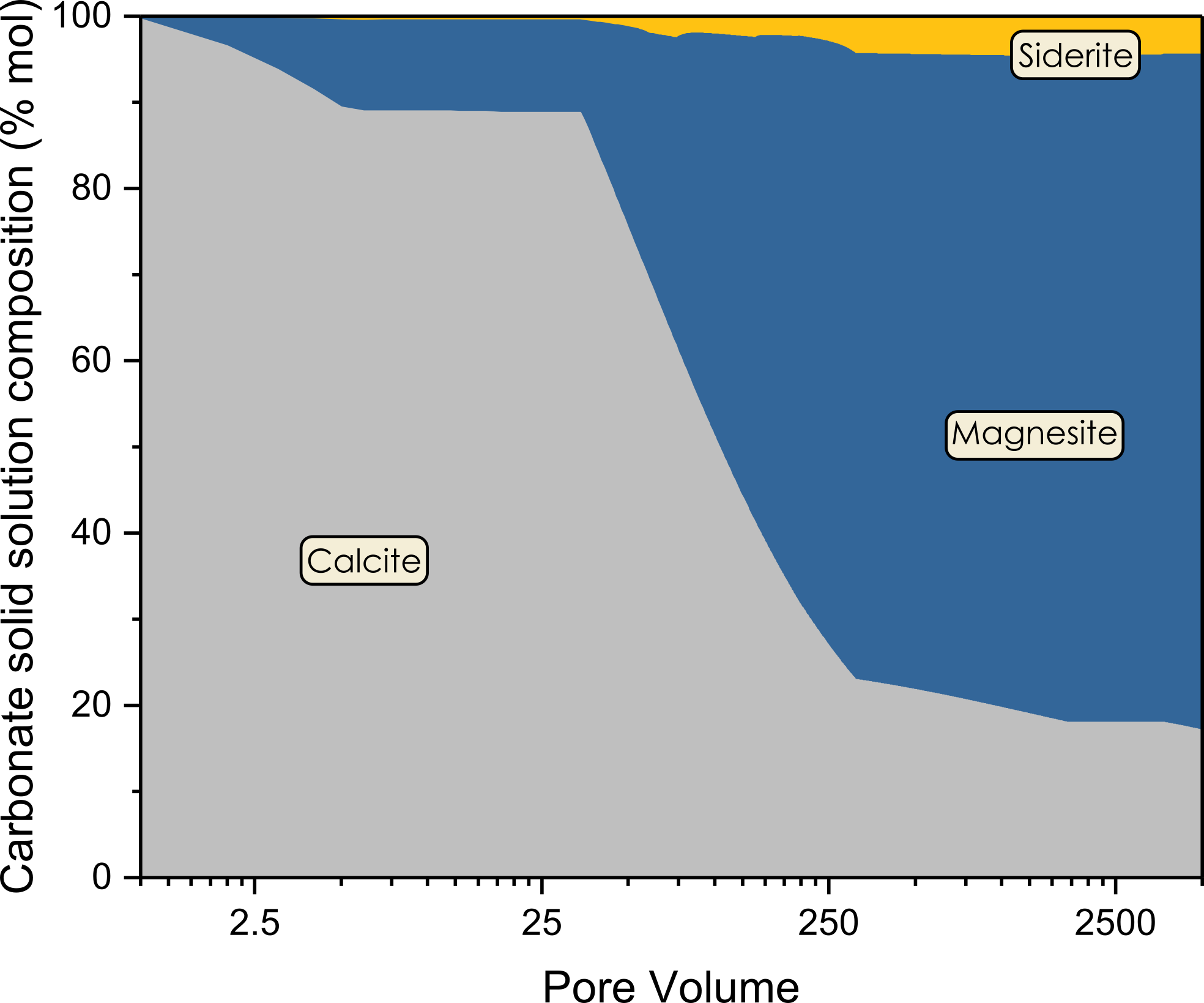}}
	
	\caption{\label{fig:assemblage_WR} Thermodynamic modeling of fluid-rock interactions with PHREEQC. \ref{fig:WR-160} and \ref{fig:WR-280}: evolution of the mineral assemblage with the W/R ratio at 160\textcelsius\ and 280\textcelsius\ respectively.  \ref{fig:assemblage_serp_160} and \ref{fig:assemblage_serp_280}: evolution of the modal proportion of the precipitated serpentine phase. \ref{fig:assemblage_increment_carbo-160} and \ref{fig:assemblage_increment_carbo-280}: evolution of the modal proportion of the carbonate phase.	The dashed line corresponds to total injected pore volume at the end of the experiments. One pore volume corresponds to W/R ratio=0.04 considering a 10\% porosity}
	\end{figure}

The equilibrium assemblage as a function of the injected pore volume (PV) is represented on Figure \ref{fig:assemblage_WR}. Results show that a serpentine/carbonate/clay assemblage coexists until 250PV  at both 160\textcelsius\ and 280\textcelsius. Further injection of the carbonate solution at 160\textcelsius\ leads to a small talc window between 250 and 500PV before the silica becomes incorporated into quartz while the rest of the divalent cations get included in the carbonate phase. The assemblage at 280\textcelsius\ follows the same  pattern with a slightly larger talc window. However beyond 3600PV quartz disappears, the silica getting incorporated into the clay phase (saponite). This relative importance of the clay fraction in the mineral assemblage is due to the high amount of aluminum in the primary minerals (pyroxenes). The Al content could potentially be even higher if the spinel phase which was not considered in the simulation was also reacting. Interestingly, the initial serpentine (Mg\#=0.9) is stable for a large range of injected PV for both temperatures while a neoformed, more iron-rich serpentine (Mg\#$\approx$0.6-0.7) is also found to be stable (Figs. \ref{fig:WR-160} and \ref{fig:WR-280}). At 160\textcelsius\ within the quartz/carbonate domain, a very small amount (<1\%mol) of serpentine is still present as pure amesite \ce{Mg2Al(AlSiO5)(OH4)} following the release of Al from pyroxene dissolution and the absence of other Al-bearing phases. Some amphiboles are also expected to form at the beginning of the injection (PV<25) as well as chlorite, while brucite is never stable due to the high amount of pyroxenes and thus the high silica activity. In terms of iron oxides, the simulations at 280\textcelsius\ suggests an increasing precipitation of magnetite during the whole process. At 160\textcelsius\, magnetite also occurs for most of the simulation albeit in smaller amounts. However, at 160\textcelsius, hematite replaces magnetite beyond 400PV while hematite only appears in HTE after 1700PV. Finally, the composition of the carbonate phase evolves from almost pure calcite at the beginning of the injection (from the recrystallization of aragonite) to a more complex solid solution incorporating Fe and Mg for increasing degrees of alteration. Aragonite is never stable in the simulations. 

\subsection{Reactive percolation experiments}
A summary of the settings for both experiments is reported on Table \ref{tab:summary_exp_perco}.

		\subsubsection{Permeability evolution}
\begin{figure}
	\centering
	\includegraphics[width=0.45\textwidth]{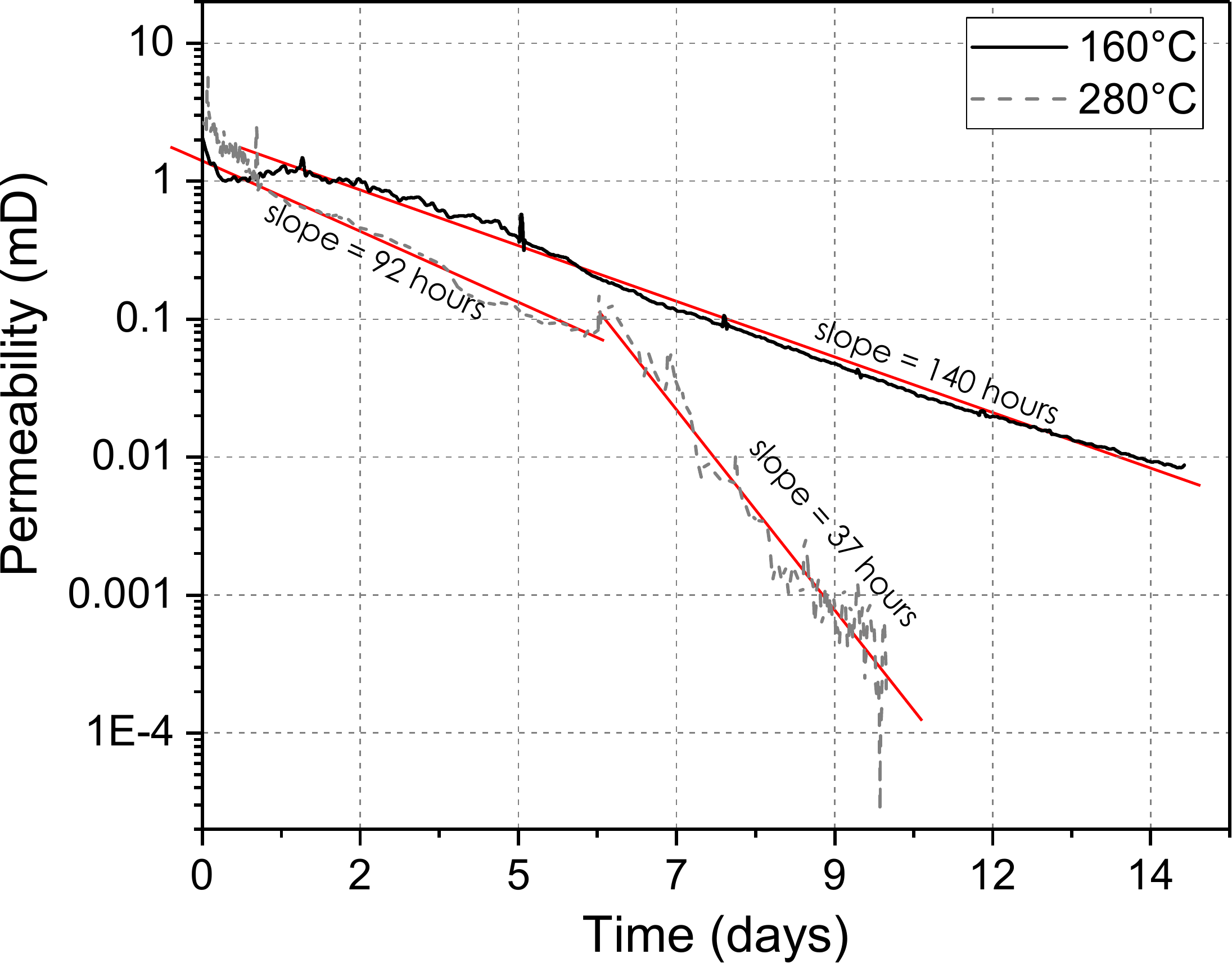}
	\caption{\label{fig:evol_permea}Variation of the permeability as a function of time for the two reactive percolation experiments}
\end{figure}

The evolution of permeability for both experiments is shown on Figure \ref{fig:evol_permea}. A global decrease is observed throughout the two experiments until cores became virtually impermeable. The decrease is in both cases linear on a log-plot (i.e. $\kappa/\kappa_0=\exp(-t/\tau)$ with $\kappa_0$ the initial permeability, $t$ the time and $\tau$ the characteristic time of decay). This exponential decrease of permeability over time is common in the description of fractured porous media subjected to dissolution-precipitation and fracture filling. Such behavior was for example described for granite under hydrothermal conditions \citep{Morrow2001}. Here, the decrease is made of a single step for LTE (characteristic time of 140 hours) and of two steps for HTE with a characteristic time of 92 hours during the first 5 days and of 37 hours thereafter. At the very beginning of LTE, the permeability shows a quick drop from 2 mD to 1mD, followed by a slight increase back to 1.5 mD before decreasing steadily. It is hard to assess if this behavior is an artifact due to the stabilization of the flow rate or if it is a consequence of some chemical or physical (fracture closing) mechanism. 

\begin{table}
\centering 
	\begin{small}
	\begin{tabular}{lcc}
	\cmidrule[1pt]{2-3}

	&	LTE	&	HTE \\
\midrule
Temperature (\textcelsius) & 160 & 280 \\
\midrule
Duration (days)	&	14 & 10 \\
\midrule
Flow rate (ml/hr) & \multicolumn{2}{c}{$\approx$0.6} 	 \\
\midrule
Fluid composition & \multicolumn{2}{c}{2\%wt NaCl, 5\%wt \ce{NaHCO3}} \\
\midrule
Core length (cm) & 3.9 & 4.1 \\
\midrule
Core diameter (mm) & \multicolumn{2}{c}{5.6} \\
\midrule
Initial mineral \mbox{composition} & \multicolumn{2}{c}{Ol, Opx, Cpx, Serp, Ara, Fe-ox, Spinel} \\
\midrule
Identified neoformed Minerals & Ara(?), Serp, Mg-\ce{CaCO3} & Serp, Cal, Hem \\
\bottomrule
	\end{tabular}
		\end{small}
	\caption{\label{tab:summary_exp_perco} Summary of the settings for the two percolation experiments}
\end{table}

		\subsubsection{160\textcelsius\ experiment (LTE)}
\begin{figure}
	\centering \includegraphics[width=\textwidth]{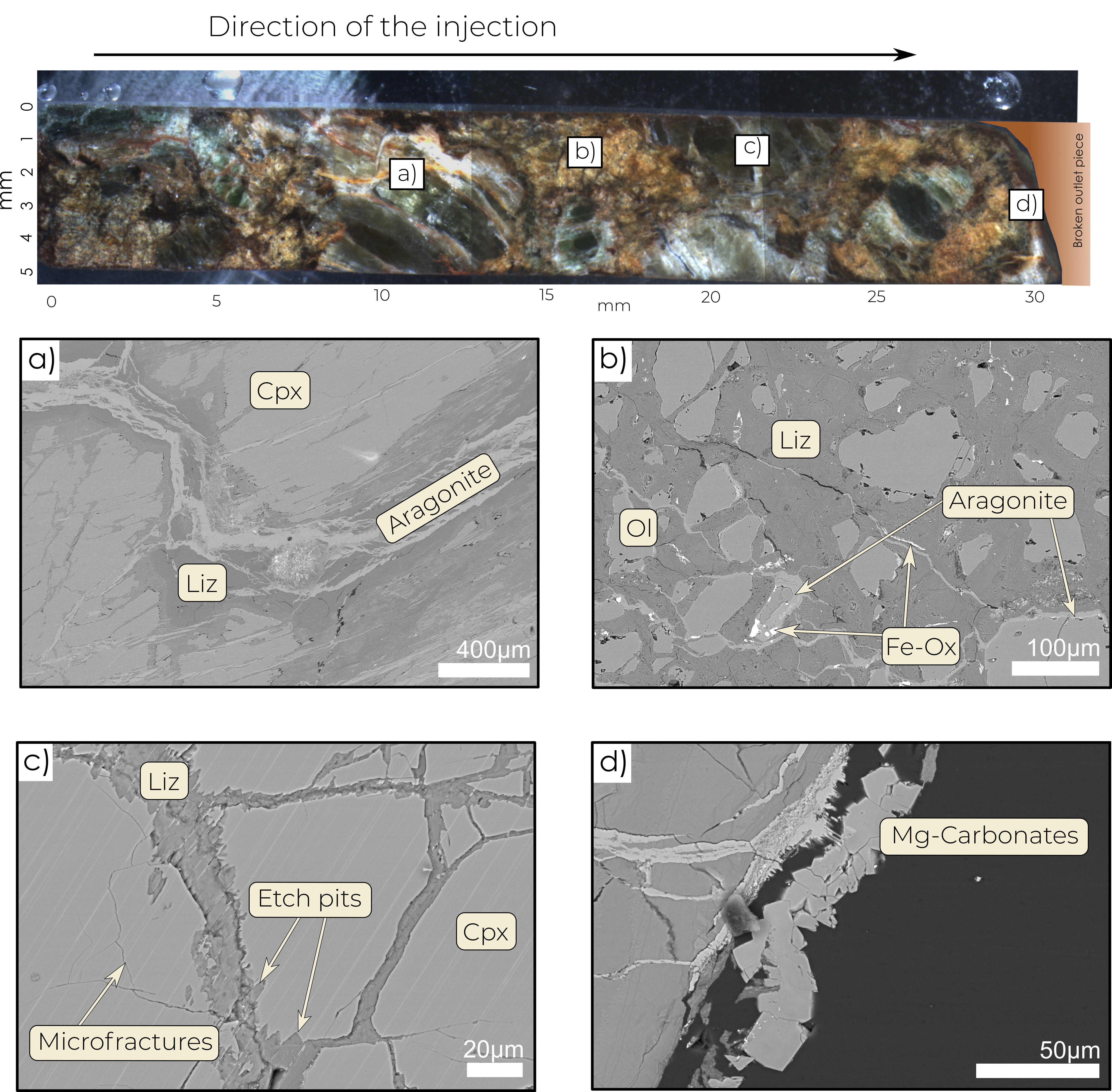}
	\caption{\label{fig:Manip5} Different views of the main part of the core reacted at 160\textcelsius. The optical view highlights the yellowish serpentine matrix and the green to dark green pyroxenes (top). SEM images a) to d) represent respectively a large carbonate vein (aragonite) through a cpx, an olivine-rich zone with carbonate rimming and veinlets, an altered clinopyroxene with microfractures originating from etch pits and probably linked to reaction-induced fracturing, and finally Mg-rich carbonates at the broken end. Due to the little alteration of the core, it is difficult to assess if these features were originally in the starting core or are a consequence of the experiment.}
\end{figure}

For this experiment, no fluid was sampled and so no information on hydrogen generation is available. Below, we concentrate on the mineralogical transformations resulting from fluid percolation. The reacted core was recovered in two pieces, the longest comprising almost the whole core, and the smallest, a few mm long, corresponding to the top end (fluid outlet). The presence of a nearly continuous layer of large ($\approx$ 20$\mu$m) euhedral carbonate crystals at the broken end of the larger piece (Fig. \ref{fig:Manip5}d) suggests that breaking did not occur during recovery but rather occurred initially during mounting of the core in its Au jacket. These carbonate crystals have a high Mg content (up to 30wt\% MgO).
	
The recovered core presents a distinct color from the initial green/brown protolith (Fig. \ref{fig:Manip5} left), appearing light brown, and even yellowish. Pyroxenes retain their green/dark green color and, so, they stand out the most clearly, while the other phases (serpentine matrix+olivine relicts) were modified. Microscopic examination shows that the olivine relicts did not significantly change in number and shape compared to the initial protolith. However, while both green and brown to orange olivine-bearing zones were initially present, the reacted core is more homogeneously colored. Carbonate layers (identified as aragonite by Raman spectroscopy) are found around most olivine relicts and aragonite veinlets connect the olivine crystals (Fig. \ref{fig:Manip5}b). In other words, the textural attributes specific of the brown to orange part in the protolith have developed and extended in the reacted core. Pyroxenes seem more altered presenting visible dissolution features, such as ubiquitous saw-tooth patterns as well as networks of microfractures originating from etch-pits (Fig. \ref{fig:Manip5}c). However, these features are difficult to attribute to the experiments with certainty because of the initial heterogeneity of the starting protolith. 

In general, aragonite veins appear yellow to orange under carthodoluminescence, with some greener and redder areas in a very similar fashion to the original protolith. Interestingly, the Raman spectra of aragonite display some variability suggesting small compositional variations consistent with the color differences seen in cathodoluminescence as well as $\mu$XRF, which shows a Mg enrichment towards the border of the veins probably due to the higher Mg activity next to the serpentine. The largest vein, which crosscuts the whole core (Fig. \ref{fig:Manip5}a) presents on the walls the same rusty color as the large veins in the initial protolith (Fig. \ref{fig:serp_init}e). This rusty zone is again composed of an intimate mixture of Ca- and Mg-carbonates and iron oxides. Since the carbonate phase is a almost exclusively aragonite, it is difficult to assess if any additional aragonite precipitation took place during this experiment. In particular, the strong similarities between cathodoluminescence images of the starting material and the LTE core (Figs. \ref{fig:cathodo_image}a,b,c) suggest that little to no precipitation occurred and the overall aragonite amount did not drastically change. Nevertheless, the Mg enrichment towards the veins walls could be the sign of extra precipitation of Mg-richer aragonite during the experiment.

		\subsubsection{280\textcelsius\ experiment (HTE)}
		
\begin{figure}
		\centering \includegraphics[width=0.8\textwidth]{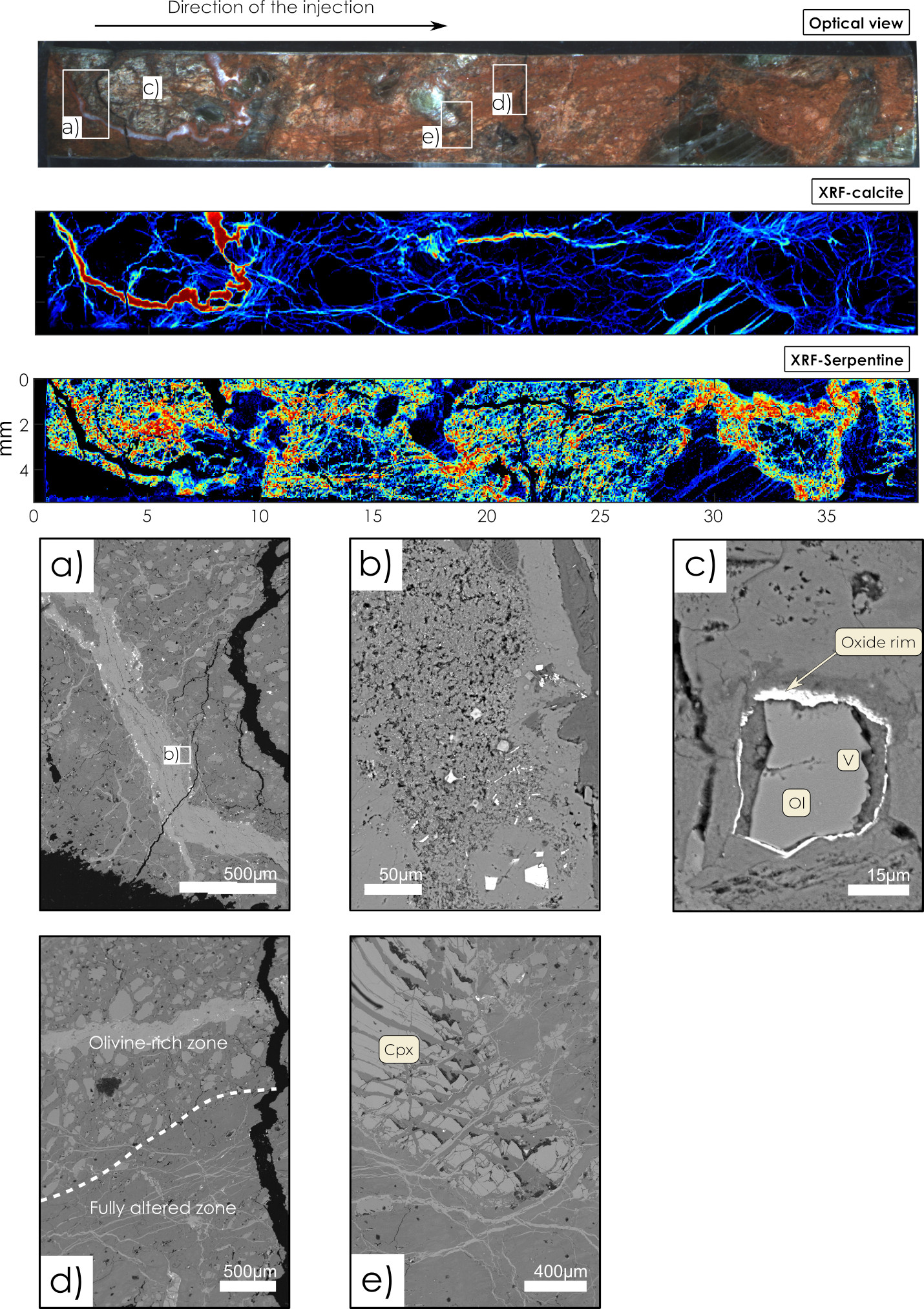}
		\caption{\label{fig:Manip6} Different views of the main part of the core reacted at 280\textcelsius. The optical view highlights the brick-red serpentine matrix and the green to dark green pyroxenes (top panel). The two $\mu$XRF panels show the difference between the calcite phase associated with oriented veins and the serpentine more broadly located and without any preferential orientation. Images and b) represent the wide calcite vein next to the inlet. Image c) pictures a relict olivine of the olivine-rich zone near the inlet with the magnetite rim. Image d) shows the juxtaposition of a fully altered zone (left) with thin veinlets and no olivines, and a less altered zone with relicts olivine and a large vein. The velocity of the flow in the large vein explains the lack of lateral alteration. Finally image e) depicts a heavily altered clinopyroxene with alteraction along the cleavage and fractures associated with reduced carbon (zoom shown in image f).} 
	\end{figure}

	\paragraph{Petrological observations}
The core after 10 days presents a striking brick-red color, in which both ortho- and clinopyroxenes appear clearly reacted, along cleavage planes as well as new fractures (e.g. Fig. \ref{fig:Manip6}e). The color is attributed to the presence of Fe(III) minerals, mainly hematite since pure cronstedtite, the Fe(III)-bearing end-member of the serpentine family, is black \citep{Hybler2016}. However, no hematite crystals could be detected by SEM suggesting that they only occur as very small crystallites. The brick-red coloring is not homogeneous in the whole sample and a gradient from fully brick-red to pink and even gray can be noted. SEM imaging reveal that this color gradient is directly linked to the presence in variable amounts of residual olivines. Zones with the strongest red color completely lack olivine and only consist of serpentine matrix plus carbonate veinlets (Fig. \ref{fig:Manip6}d, bottom), indicative of a high degree of alteration. Conversely, the least colored zones contain olivines (the lighter the color, the more the olivine relicts) and in particular the area next to the inlet. This part of the sample appears to have been barely modified during the experiment and it shows features strikingly similar to green zones described in the starting serpentinite. Yet, the pre-existing magnetite rim (Fig. \ref{fig:serp_init}c) is now found included in the serpentine matrix at some distance away from olivine suggesting a volume reduction of the olivine cores (Fig. \ref{fig:Manip6}c).

Near the inlet, a wide ($\approx$0.5mm) carbonate vein wraps the untransformed olivine-rich zone. It shows the same characteristics (in particular the rust colored border with porous texture and presence of Fe oxides and Mg-rich carbonates) as aragonite veins found both in the LTE and in the starting serpentinite (Fig. \ref{fig:Manip6}a and b). This vein marks the infilling of a large fracture visibly open and empty in the starting core. The rest of the carbonates occur throughout the sample as a very dense network of veins and veinlets developed parallel to the flow direction and identified as calcite with Raman spectroscopy. Cathodoluminescence shows that the network of calcite is significantly denser than the aragonite network of the starting material and presents a very homogeneous and extremely bright red color, which contrasts with the less luminous color gradients of the aragonite veins in the LTE and the initial serpentinite (Fig. \ref{fig:cathodo_image}). The only parts appearing dark in cathodoluminescence are the relicts of pyroxenes. There is an inverse correlation between the density of the veinlets network and the presence of olivine, with the redder zones containing numerous small veinlets while the less altered zones present either few veinlets or just a large vein cutting across (left and right area of Fig.\ref{fig:Manip6}d). As highlighted by the $\mu$XRF imaging, the calcite network is only distributed within fractures and veins and appears aligned with the flow direction. On the contrary, serpentine, does not show any preferential orientation. Finally, spot analysis with electron microprobe (see supplementary data) did not allow to identify unambiguously the neoformed serpentine, as all serpentine compositions which were analyzed on HTE did not statistically deviate from the serpentine composition in the starting material. 

\paragraph{Chemical analysis}
Five aliquots of outlet fluid were sampled at different and increasing time intervals over the entire experiment and analyzed for major cations and anions as well as for dissolved hydrogen. The data yield the average concentrations of the effluent between two sampling events. Additionally the outlet solution is being diluted when reaching the pump as the outlet pump is refilled with some deionized water after each sampling. Data were rescaled by this dilution factor to obtain the actual outlet concentrations. Results are presented in table \ref{tab:Results_Manip6}, with the first number the measured concentration and the second, the rescaled concentration.

Chloride concentration is mostly conserved during the experiment. The small variations observed could possibly reflect the precipitation of halite, of which a few crystals were detected on SEM, as well as iowaite $\ce{Mg_6Fe^{3+}_2(OH)_{16}Cl_2* 4H_2O}$ even if this mineral was not observed. Outlet Na and DIC concentrations appear systematically lower than the injected amount, suggesting some uptake by the core. Magnesium and calcium show very similar evolution with concentrations below detection in the middle of the experiment and two spikes, one at the very beginning and the other at the very end of the experiment. We note that the inlet solution contains some Ca (0.39 mg/L) from imperfectly exchanged deionized water. This small Ca background concentration actually  facilitates mass balance comparisons. In particular, the total absence of Ca in the outlet fluid following the early spike stresses that this element has been completely trapped in the core. The dissolved hydrogen follows a trend similar to Ca and Mg with a first spike at 100$\mu$mol/kgw on the first day, followed by steady concentrations of around 30-50 $\mu$mol/kgw in the following days and finally a strong increase up to 162 $\mu$mol/kgw in the last days. 

The large uptake of DIC is supported by the very dense calcite network visible in Figure \ref{fig:Manip6}. It is also possible that part of the injected DIC was reduced to organic molecules or even graphite, however not evidence of this reaction was found. In parallel, the core has also taken up an unexpected amount of sodium (5.6 mmol) which could not be related to any visible mineral. We can easily rule out any precipitation of sodium-bearing mineral in the tubing or in the pump, as the only mineral which could precipitate is halite and the uptake of chloride is two orders of magnitude smaller (0.2 mmol). A combination of three options remain to explain the sodium behavior: clay precipitation, Na-bearing carbonate precipitation or Na-adsorption. The first option is supported by PHREEQC simulations with clay phase composed of saponite and vermiculite. However, the maximum quantity of Na mineralized in the simulations is around 0.1 mmol at an injected pore volume of 40 which is insufficient to explain the Na uptake. The other option is the precipitation of Na-bearing carbonates, such as the rare mineral eitelite (\ce{Na2Mg(CO3)2}). Finally the last possibility is the adsorption of Na on serpentine. It is however difficult to assess the relative importance of these phenomena with certainty.

\begin{table}
\begin{scriptsize}
	\begin{tabular}{cccccccccccccc}
	\toprule
Sample & Elapsed time &  \multicolumn{2}{c}{Cl} & \multicolumn{2}{c}{DIC} & \multicolumn{2}{c}{Na} & \multicolumn{2}{c}{Ca} & \multicolumn{2}{c}{Mg} &\multicolumn{2}{c}{\ce{H2}} \\
		\midrule
		& days		 	& mg/L & mol/L & mg/L & mol/L & mg/L & mol/L & mg/L & $\mu$mol/L & mg/L & $\mu$mol/L & \multicolumn{2}{c}{$\mu$mol/kgw} \\
		\midrule
Inlet solution	& 	- 	  & 12,160 & 0.342 & 7,190 & 0.599 & 21,920 & 0.954 & 0.39 & 9.7 & 0 & 0  & 0 & 0 \\
		\midrule
Sample  1		& 	1		&  - & - & - & - & 10,453 & 0.792 & 1.23 & 53.6 & 0.13 & 9.3 & 44 & 76 \\
\midrule
Sample 2 		& 	2	&  7,090 & 0.270 &  4,650 & 0.523 & 14,680 & 0.863 & 0.22 & 7.42 & 0.12 & 6.67 & 20 & 27 \\
\midrule 
Sample 3		& 	3	&  8,640 & 0.299 &  5,430 & 0.556  & 16,530 & 0.884 & 0    & 0	  &  0   & 0    & 26 & 32 \\
\midrule 
Sample 4		&	6		& 8,740 & 0.291 &  5,660 & 0.557 & 17,923 & 0.921 & 0    & 0    &  0   & 0    & 36 & 42\\
\midrule 
Sample 5		& 	9		& 7,540 & 0.262 &  5,700 & 0.587 & 17,216 & 0.926 & 0.41 & 12.6 & 1.28 & 65.1 & 125 & 154 \\
\bottomrule

\end{tabular}	
	\end{scriptsize}
\caption{\label{tab:Results_Manip6} Chemical analysis for samples from the reactive percolation experiment at 280\textcelsius.}		
\end{table}

\begin{figure}
	\centering \includegraphics[width=0.9\textwidth]{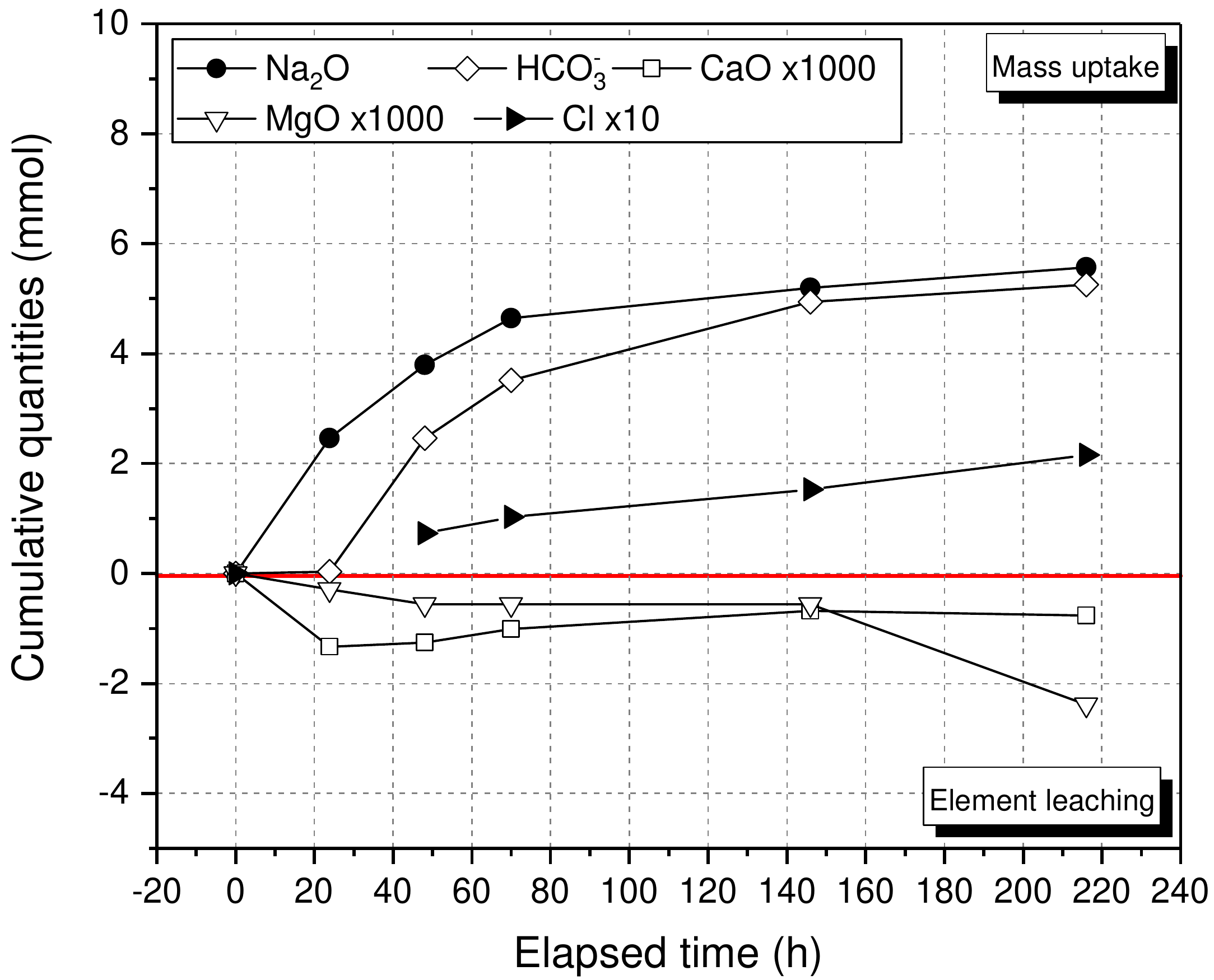}
	 \caption{\label{fig:massbalance} Cumulative mass balance for the measured oxides for HTE}
\end{figure}

\section{Discussion}

\subsection{Mineralogical evolution and pore size distribution}
Despite very similar experimental conditions, except for the experimental temperature, HTE and LTE present fundamentally different mineralogical and petrophysical behaviors. Indeed, the LTE presents little evolution from the starting material, with relicts olivine and pyroxenes embedded in a serpentine matrix through which run veins and veinlets of aragonite. However, despite the absence of obvious modifications from the starting material mineralogy, the fast decrease in permeability of LTE is still evidence of a precipitation leading to the clogging of the permeability, either from the serpentinization of olivine and pyroxenes, or from the precipitation of extra aragonite in the open porosity (such as the Mg-enriched aragonite at the interface between serpentine matrix and aragonite veins). The HTE presents on the contrary, not only a complete replacement of the aragonite network by calcite, but also a significant increase in the number and density of calcite veins throughout the core.  The permeability decrease is significantly faster and is consistent with the densification of the calcite network. Since the starting material was initially relatively altered, it is quite difficult to quantify the alteration of the silicate minerals, but the strong alteration features (olivine-free zones on Figure \ref{fig:Manip6}d, or heavily altered cpx \ref{fig:Manip6}e), which have no equivalent on the starting material, suggest nevertheless that silicates where deeply altered during the experiment. The chemical modification of the mineral assemblage of HTE is also evidenced from chemical analysis of the outlet fluid with a leaching of Mg and Ca from the dissolution of the starting minerals, a production of hydrogen through Fe oxidation, and an uptake of \ce{CO2} feeding the calcite precipitation. 

In the literature, the reactivity of ultramafic minerals to carbon-rich fluids has been extensively studied both in natural environments \citep{Hansen2005,Falk2015} and in batch experiments \citep{OConnor2005,Hovelmann2012}. The transformation of ultramafic rocks subjected to the percolation of carbon-rich fluids is usually described as a gradual evolution towards first a serpentine/carbonate assemblage at low degrees of alteration, then to a talc/carbonate assemblage (soapstone) following the increase of the Si/Mg ratio of the fluid due to the continuous dissolution of the silicate minerals and the uptake of Mg by the precipitating carbonates \citep{Beinlich2012}. At higher degrees of alteration, the further increase in Si/Mg ratio of the fluid leads to a final assemblage quartz/carbonate called listvenite, where all the silica from the primary minerals is taken up by the quartz phase, while the divalent cations are incorporated in the carbonate phase. According to \citet{Menzel2018}, this evolution can occur without any other input than the \ce{CO2} content of the infiltrating fluid. This succession is observable in numerous natural settings \citep{Falk2015,Menzel2018,Power2013a} and is perfectly consistent with our numerical results (Fig. \ref{fig:assemblage_WR}) but is however not universal and some phases (e.g. talc) may be omitted depending on the percolating fluid composition and its  velocity \citep{Hinsken2017}.

The final composition of both LTE and HTE -- i.e. mostly serpentine + \ce{CaCO3} (aragonite for LTE and calcite for HTE), relictual pyroxenes and olivine, iron oxides, absence of talc, quartz, or clay minerals -- is characteristic of low alteration levels and corresponds to an injected pore volume up to 250 in the simulations (Fig. \ref{fig:assemblage_WR}). However, the total injected pore volume of the experiments is actually one order of magnitude larger. PHREEQC simulations are indeed performed at equilibrium thus considering an infinite reaction rate at each step, or an infinitely slow flow rate (a case coined transport-limited as the extent of reaction does not depend on the reaction kinetics but is only limited by the capacity of the system to transport chemical species to and from the reactive surfaces). In the presented experiments, the actual behavior corresponds to an intermediate case with different and finite reaction rates for each mineral reaction. As a result, the system requires more fluid to achieve the same degree of alteration as the simulations, since the percolating fluid does not have enough time to fully react before exiting the core, limiting effectively the actual input in terms of dissolved \ce{CO2}. This relationship between chemical reaction and mass transfer by advection is scaled by the \Da\ number, $Da = \frac{k}{vc}$, with $k$ the intrinsic reaction rate in mol.m$^{-2}$.s$^{-1}$, $v$ the fluid velocity (from Darcy's law) and $c$ the solubility in pure water of the considered mineral. Large Da correspond to situations where the reaction occurs more rapidly than advection, either because of very fast kinetics or very slow flow rates similarly to PHREEQC simulations. This leads to a system at chemical equilibrium for every point in space and time. On the contrary, small Da correspond to situations where the reaction is much slower than the fluid velocity and the system remains permanently out of equilibrium (reaction-limited regime). 

Since Da depends on both the intrinsic kinetics of the considered mineral as well as the fluid velocity, we can identify in our experiments three different zones with three different behaviors, following the three classes of pores identified by Mercury Intrusion Porosimetry (i.e. >62$\mu$m, [1-62] $\mu$m and <1$\mu$m).

\begin{figure}
	\centering \includegraphics[width=\textwidth]{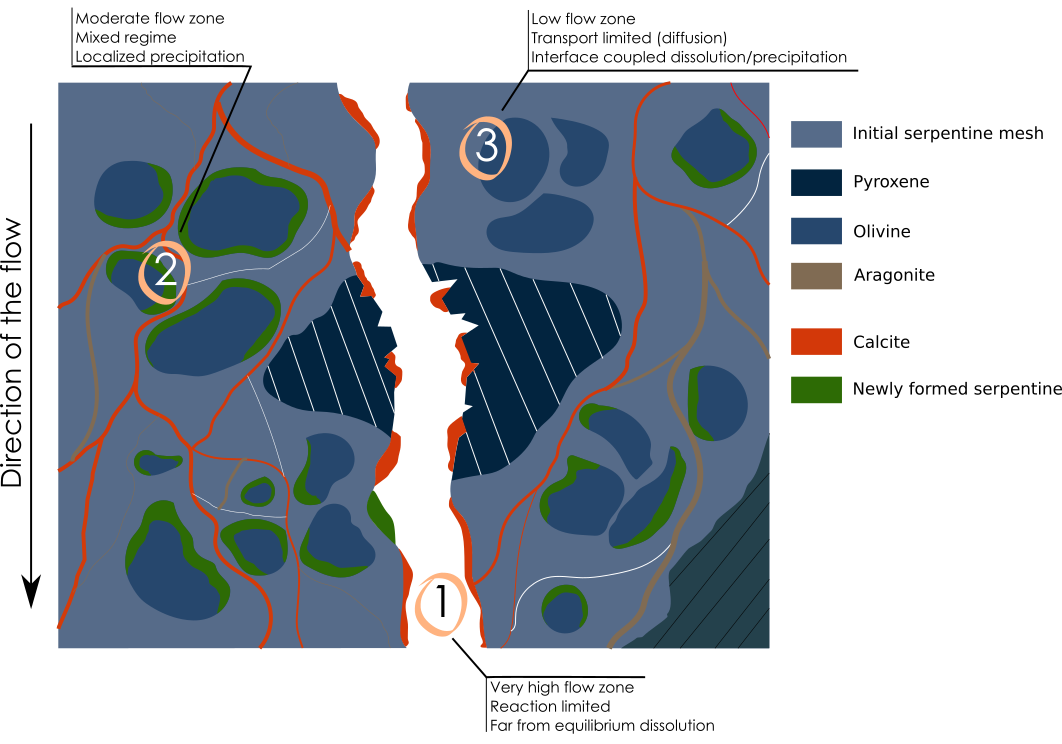}
	\caption{\label{fig:schema_final}Schematics summarizing the overall behavior during the reactive percolation experiments at 280\textcelsius. The behavior can be divided into three main zones, with the larger fractures seeing most of the flow, the smaller fracture network seeing an intermediate flow rate while the porous serpentine matrix sees little to no flow. This has consequences on the location of the dissolution and precipitation of the different minerals. }
\end{figure}

\begin{itemize}
	\item \textbf{Large veins} In the larger veins, which correspond to the larger pore network (Zone 1 on Figure \ref{fig:schema_final}), dissolution of the primary minerals will be homogeneous over the whole length of the system and can be considered as far-from-equilibrium since these zones concentrate the largest flow. The system will be reaction-limited (Da $\ll$ 1) as the quantity dissolved per unit time is only linked to the intrinsic dissolution rate of each mineral and their reactive surface area. For slow precipitating minerals (i.e. serpentine), precipitation will occur far from the inlet and most of the solutes will actually exit the system without any actual precipitation (net mass loss as confirmed by the Mg leaching on HTE). On the other hand, fast precipitating minerals (i.e. calcite and potentially hematite) will still be able to precipitate in these fractures (the only minerals with Da $\gg$ 1, see Supplementary Information). The fact that some Ca is still leached from the core (albeit in smaller quantities than Mg) shows that despite the very fast kinetics of calcite precipitation, the conversion of aragonite to calcite is not complete. 
	
	\item \textbf{Small veinlets} For the second class of pores (Zone 2 on Figure \ref{fig:schema_final}), on the contrary, the fluid velocity is significantly lower leading to a mixed- to transport-limited regime (i.e. Da $\gg$ 1 for all minerals except cpx and magnesite) and a more localized reactivity. This means that the dissolution and precipitation of minerals will rather occur locally as an interface controlled process \citep{Ruiz-Agudo2014}, close to equilibrium, with the neoformed minerals precipitating at the same location than dissolution occurred. This situation can lead to the pseudomorphic mineral replacement by the synchronization of the reaction fronts \citep{Putnis2009,Kondratiuk2017}. Moreover, because of the limited flow, overall little \ce{CO2} is brought to these areas meaning that little carbonation actually occurs, and only in the largest veinlets. 
	
	\item \textbf{Small pores and fractures (<1$\mu$m)} For the third class of pores (<1$\mu$m), the velocity is almost negligible and the solute transport is exclusively controlled by diffusion. This leads to a full transport-limited regime and thus very slow reaction rates. Similarly to the small veinlets, due to the extremely small flow rate, no \ce{CO2} is able to reach these zones and so no carbonation occurs, allowing only serpentinization reactions to take place. 
\end{itemize}

At the core size, an interesting example of the relationship between kinetics and pore network can be observed on Figure \ref{fig:Manip6}d, highlighting the fundamental importance of diffusion and advection. The top part of the image presents a large vein cutting through an olivine-rich area, while the bottom part presents smaller veins and almost no olivine remaining. In the top part, the velocity in the vein is large and the ratio between advection and diffusion (Péclet number) is much larger than 1, implying an advection-dominated regime, while the rest of the surrounding matrix is diffusion-dominated. Since there is no advection in the matrix the replacement reaction progresses very slowly, while dissolution/precipitation occurs more rapidly in the large vein due to the fast advection. On the other hand, for the bottom part of the image, the alteration is much more homogeneous as numerous fractures and veinlets percolate through the serpentine matrix. The density of these veinlets allows for a faster elimination of the reaction products as well as an increased input of dissolved \ce{CO2}, accelerating olivine dissolution, consistent with the complete absence of olivine relicts in this area. This type of observation can be extended to the whole core: wherever there is an extended network of small veins, the system is strongly altered, while large veins seem to traverse the system without much interactions. Another prime example of this behavior is the inlet of HTE, with the very large vein of carbonate right next to a zone which was barely modified. During the experiment little to no fluid at all flowed in this particular zone and thus little to no \ce{CO2} because all the fluid was funneled in the larger fracture. As a result, only pure serpentinization occurred leading to the clear "green"-type textural features. 

\subsubsection{Carbonates precipitation}

Aragonite presents the fastest dissolution kinetics of all the primary minerals at both temperatures. In the absence of any other chemical reaction, this would mean that the solution percolating through the sample will get quickly saturated and dissolution would occur only next to the inlet with a dissolution front progressing in the sample following the depletion of aragonite (Da $\gg$ 1). However, dissolution of aragonite is coupled with the precipitation of calcite, which is more thermodyamically stable at both temperatures and is just as fast to react as aragonite. As a result, as soon as supersaturation is generated by aragonite dissolution, calcite should immediately precipitate. The consequence is that aragonite will never be at equilibrium with the fluid and will keep dissolving as long as calcite is able to precipitate. 

In HTE, aragonite is completely replaced as expected by a very chemically-homogeneous calcite, over the whole length of the core. However, in LTE aragonite remains preserved and no calcite was identified by Raman spectroscopy. Since thermodynamics predicts the total replacement of aragonite by calcite for both LTE and HTE, this means that calcite precipitation was slowed or even inhibited, preventing the further dissolution of aragonite. This inhibition can be explained by the effect of dissolved Mg on calcite precipitation kinetics. Indeed, calcite precipitation is known to be strongly slowed down or even inhibited in the presence of magnesium ions \citep{Zhang2000}. This is one of the alleged reasons for the existence of aragonite in marine serpentinites out of its usual high pressure domain of stability \citep{Bonatti1980}. Thus calcite precipitation could have been slowed enough in the LTE to explain its absence, leading to an aragonite-saturated solution in the whole core due to its fast dissolution kinetics. The consequence is that the carbonate network was untouched in most of the core, with only the inlet being subjected to aragonite dissolution. Interestingly, the carbonate veins next to the inlet appear extinct under cathodoluminescence suggesting an evolution of the composition from the rest of the core and potentially corroborating the hypothesis. In this context, the complete replacement of aragonite by calcite in HTE could be explained by a less effective suppression of calcite precipitation by Mg$^{2+}$ at a higher temperature, although calcite precipitation is probably still slower than in the absence of Mg since a small amount of Ca was still able to exit the core.

Similarly, the complete absence of magnesite can be explained by two concomitant reasons: first, magnesite is known to have very slow precipitation kinetics \citep{Saldi2012}. This is for example consistent with the observations of \citet{Andreani2009} and \citet{Peuble2015a} who observed that magnesite was preferentially precipitating in zones of low flow, where the fluid resides long enough for magnesite to precipitate (Da $\gg$ 1 due to smaller $v$). This is consistent with the detection of Mg-enriched carbonates at the rim of the large veins or of the large Mg-rich crystals from  Figure \ref{fig:Manip5}b which appeared due to the exceptional slowing of the fluid in this large artificial fracture. Considering the large difference in precipitation kinetics between calcite and magnesite, dissolved inorganic carbon will be preferentially incorporated in calcite rather than in magnesite, inhibiting the precipitation of the latter. Secondly, in our experimental conditions, serpentine precipitates faster than magnesite and so the competition for Mg should also favor the latter. The complete absence of magnesite is then due the favored precipitation of calcite in high DIC zones and of serpentine in high silica zones. Magnesite or Mg-rich carbonate solid solution (even potentially dolomite) is however expected to occur at high degrees of alteration as highlighted by PHREEQC simulations since the continuous supply of DIC forces the dissolution of serpentine and the release of Mg, leading to the progressive incorporation of Mg in the carbonate solid solution.

\subsubsection{Iron oxides}
Numerical simulations predicts the precipitation of hematite after 1700PV at 280\textcelsius\ which is consistent with the brick-red coloring of the core. This precipitation of hematite, which has already been observed in similar situations (e.g. \citet{Godard2013}), can be explained by a combined action of hydrogen removal by the continuous flow of solution and the constant input of dissolved oxygen from the injected fresh solution (a calculated concentration of 0.226 mmol/kgw). The same simulation without dissolved oxygen produces only magnetite even at the highest injected pore volumes consistent with natural serpentinization, where magnetite is found almost exclusively and hematite is very rare due to the anoxic nature of serpentinizing fluids.  In contrast however, hematite is also predicted in the simulations at 160\textcelsius\ but does not seem to have precipitated in the experiment. This discrepancy is easily explained by the fact that the simulations predict hematite when all the primary minerals are altered and the system reaches the quartz/carbonate assemblage, which is not the case since the alteration in LTE  corresponds rather to an injected pore volume of 250 or less due to the finite reaction kinetics. 

\subsection{Permeability evolution and comparison with the literature}
In order to interpret our permeability data, results from similar experiments in the literature are reported in table \ref{tab:literature}. The first 3 rows correspond to carbonation experiments while the next three are simple serpentinization for comparison. Carbonation experiments, even with strongly varying experimental conditions, report consistent values of the characteristic decay time of 140 and 185 hours for experiments at 160\textcelsius\ and 180\textcelsius\ (this study and \citep{Peuble2015a}). Increasing temperature seems to accelerate the decrease in permeability decay time down to a narrow range around 40 hours. Interestingly, both this study at 280\textcelsius\ and the slower flow rate in \citep{Peuble2015a} report an induction period of a few days with slower permeability decrease (respectively 92 and 80 hours). On the contrary, pure serpentinization experiments report a much slower permeability decay rate with values ranging from 2000 to 10,000 hours. 

The difference stems from the difference in precipitation kinetics between silicate (serpentine) and carbonates. In particular, for higher flow rates, the slower serpentine precipitation leads to a non isochemical behavior with most of the dissolved serpentine exiting the system instead of precipitating (Mg leaching). The stark decrease in characteristic time from the experiment of \citet{Godard2013} when the flow rate is decreased suggests an increase of precipitation and thus a faster decrease in permeability. We recalculated the \Da\ number in their experiment and obtained values of 10$^{-5}$ for olivine and 10$^{-4}$ for serpentine (the change of flow rate does not modify the order of magnitude) confirming a reaction-limited behavior but seemingly at odds with the author's interpretation of fast kinetics and a transport-limited regime. However, similarly to our experiments, it is possible that some zones of lower flow rate can have a more transport-limited regime impacting strongly the overall behavior. For these zones to appear, it would require a decrease of the Darcy velocity of 4 orders of magnitude or pores 10 times smaller (from the calculation of hydraulic resistance, considering similar surface areas), which are very likely to exist in their system. 

Additionally, during serpentinization, the largest amount of precipitation occurs in the smallest pores where the velocity is reduced enough to be closer to equilibrium. Since the hydraulic conductivity of these pores is already very small, their clogging does not impact heavily the permeability. It is only when the quantity of precipitates becomes significant in the larger pores that the overall permeability is reduced which explains the very slow clogging time in these experiments. Interestingly, the characteristic time in our experiments, for serpentine precipitation (defined as $\tau = c/(k\times s_R)$, with $s_R$ the reactive surface area) is around 5700 hours for serpentinization in the larger class of pores at 160\textcelsius. This value is of the same order of magnitude than the permeability decay time for serpentinization-only experiments showing thus clearly the control of serpentine precipitation on the evolution of permeability (similarly to \citet{Godard2013} conclusions) and thus the reaction-limited characteristics of the experiments. 

On the other hand, carbonation experiments present a much faster decay rate because of the much faster kinetics associated with carbonate precipitation. Since carbonate precipitation can occur in a broader class of pores (and in particular in the large main percolation paths), permeability is more heavily and more rapidly impacted. The characteristic time for the precipitation of magnesite in our experiments  is around 20 hours, similar to values reported for the carbonation of sintered olivine, showing again the control of the precipitating mineral on the permeability decay in case of reaction-limited experiments (as described by \citet{Peuble2015a}). Yet, permeability in HTE decays much slower than the characteristic time of aragonite to calcite transformation (of the order of the millisecond) or even the precipitation of extra calcite. This is because for HTE, the aragonite/calcite replacement is not reaction-limited (because of the fast kinetics) but transport-limited. The control on the permeability decay is then a coupling between time of residence of the fluid and replacement kinetics. Interestingly, for LTE, where the aragonite/calcite replacement is not expected to happen to a large extent, the decay time is still much faster than serpentinization suggesting that some extra carbonate precipitation occurred to some extent, through the precipitation of the co-injected Na$^+$, Ca$^{2+}$ and DIC and the presence of Mg$^{2+}$ from silicate dissolution. 

The literature often considers that precipitation of carbonates is less impactful on the permeability because of the small volume change associated compared with silicate precipitation (e.g. \citet{Beinlich2012}). Our experiments suggest the contrary as the decay of permeability is more consistent with carbonate precipitation than silicates. It is actually a well-known phenomenon that a small variation of porosity can lead to dramatic drop in permeability \citep{Muller2009} if precipitation occurs in 'strategic' locations, and even if the associated volume change  is small.

\begin{table}
\begin{scriptsize}
	\begin{tabular}{cccccc}
	\toprule
	Reference & flow rate & Characteristic time & Temperature & Core & Solution\\
			&	ml/hr	&	hr				&	\textcelsius &	&	\\
			\midrule
	\multirow{2}{*}{\citet{Peuble2015a}}	& 0.1  & 80 then 36 after 1.5 day & \multirow{2}{*}{180} & \multirow{2}{*}{sintered olivine} & \multirow{2}{*}{NaHCO3 buffered - pCO2 = 10MPa}\\
										& 1	&	40	&		&		&		 \\
	\midrule	
	\citet{Peuble2018} & 12 then 6	&	185 & 160 & sintered olivine & NaCl 1M, NaHCO3 0.4M, pCO2 = 11MPa \\	
	\midrule
	\multirow{3}{*}{\citet{Peuble2019}} & \multirow{3}{*}{0.5} & 363 then 85 after 5 days & \multirow{3}{*}{185} & \multirow{3}{*}{sintered olivine} & Volvic\textregistered\ + 10MPa$_{\ce{CO2}}$ \\
										&					&	147 then 49 after 4 days & 	&  & Volvic\textregistered\ + 1MPa$_{\ce{CO2}}$ \\
										&					&	51		&	&	& Volvic\textregistered\ + 0.1MPa$_{\ce{CO2}}$ \\ 
	\midrule[1pt]
	\multirow{2}{*}{\citet{Luhmann2017b}}	& 0.01 	&	 2200 & 150 & \multirow{2}{*}{intact dunite} & \multirow{2}{*}{Artificial seawater} \\
										&	0.01 	&	1850 (130 in the first days) & 200 & 	& \\	
	\midrule 
	\citet{Godard2013}	&	0.2 then 0.06 after 8 days	&	10300 then 330 & 160	&	sintered olivine & Artificial seawater \\
	\midrule[1pt]
	\multirow{2}{*}{This study} &	\multirow{2}{*}{$\approx$0.6} & 140 & 160 & \multirow{2}{*}{natural serpentinite} & \multirow{2}{*}{2\%wt NaCl, 5\%wt NaHCO3} \\
								&							& 92 first 6 days, 37 after & 280 & 	&	 \\
	\bottomrule
	\end{tabular}
	\end{scriptsize}
	\caption{\label{tab:literature} Characteristic decay time for permeability evolution for reactive transport experiments of pure serpentinization and carbonation of ultramafic rocks from the literature.}
	\end{table}

\subsubsection{Hydrogen production and \ce{CO2} sequestration}
Hydrogen production in mid-ocean ridge is a fundamental process in the seawater cycle through the oceanic crust as well as an important component in the deep-sea biome and ecosystem. However, estimation of the global \ce{H2} output from the mid-ocean ridges is incredibly challenging as the output does not only depend on the raw quantities of water and peridotite but also on the  coupling between transport and reactivity. The numerical simulations in this article, based on a fully transport-limited regime, reproduce quite accurately the hydrogen concentrations however it significantly overestimates the total amount produced with an expected 280 mmol compared to the total of 10 $\mu$mol measured during the experiment. While the measurement of hydrogen is probably underestimated because of losses in the outlet pump as well as during the sampling the large difference is once again due to kinetics reasons as the slower dissolution in the experiments lead to a slower release of Fe$^{2+}$ in solution and thus a slower \ce{H2} production rate. The other reason for the lower production of hydrogen is the potential reduction of DIC to carbon suggested by the large amount of \ce{CO2} retained in the core but not evidenced in the characterization. One could expect that the total \ce{H2} could get closer to the expected amount the longer the experiment lasts, but the complete clogging of the system from carbonate precipitation prevented to test this hypothesis. in the end, the total yield before clogging of HTE was only 0.8\% for hydrogen production if taking into account all the Fe$^{2+}$ in the initial core. 

Studies on oceanic serpentinization and \ce{H2} production usually invoke reaction-induced fracturing to explain sustained permeability despite the large unfavorable volume change between peridotite minerals and serpentine. However, for reaction-induced fracturing to occur, a substantial supersaturation has to be built as the force pushing minerals to precipitate should overcome the mechanical resistance of the surrounding matrix \citep{Osselin2013a,Osselin2015b}. In the case of calcite precipitation, the reaction is very fast and no high supersaturation should be allowed to build-up leading to small crystallization forces and \emph{in fine} little to no fracturation. Our experiments suggest then that carbonation is a self-limiting process which, if happening in parallel with serpentinization might strongly hinder hydrogen production by clogging the permeability and preventing deep fluid infiltration. However, one need to keep in mind that the injected solution in our experiments is extremely DIC-rich and with a pH favoring rapid precipitation of carbonates. More acidic or less concentrated fluids would lead to a slower precipitation of carbonates and thus to a smaller impact on the permeability. 

This permeability impairment due to carbonate precipitation poses a huge threat on the viability of carbon sequestration techniques relying on the reactivity of ultramafic rocks, using leached Mg and Ca as counter-cations for the precipitation of carbonates. Our experiments show that substrates containing significant amounts of natural carbonates is a potential liability as the fast reprecipitation of these carbonates depending on the thermodynamic disequilibrium induced by the percolating fluid might lead to a fast and deadly drop in injectivity. In the same vein, substrates with a high content in clinopyroxenes might not be ideal either due to the fast precipitation of Ca-bearing carbonates. This is however balanced by the extremely slow dissolution kinetics of cpx in these conditions. The paradoxical conclusion is that in order to get the best long-term yield in terms of hydrogen production and \ce{CO2} sequestration, it is better to slow down the kinetics of carbonate precipitation so that they do not precipitate in the main percolation paths, but rather in the smaller pore network, similarly to serpentine, where they would not have such a dramatic impact on permeability.

\begin{figure}
	\centering  \includegraphics[width=\textwidth]{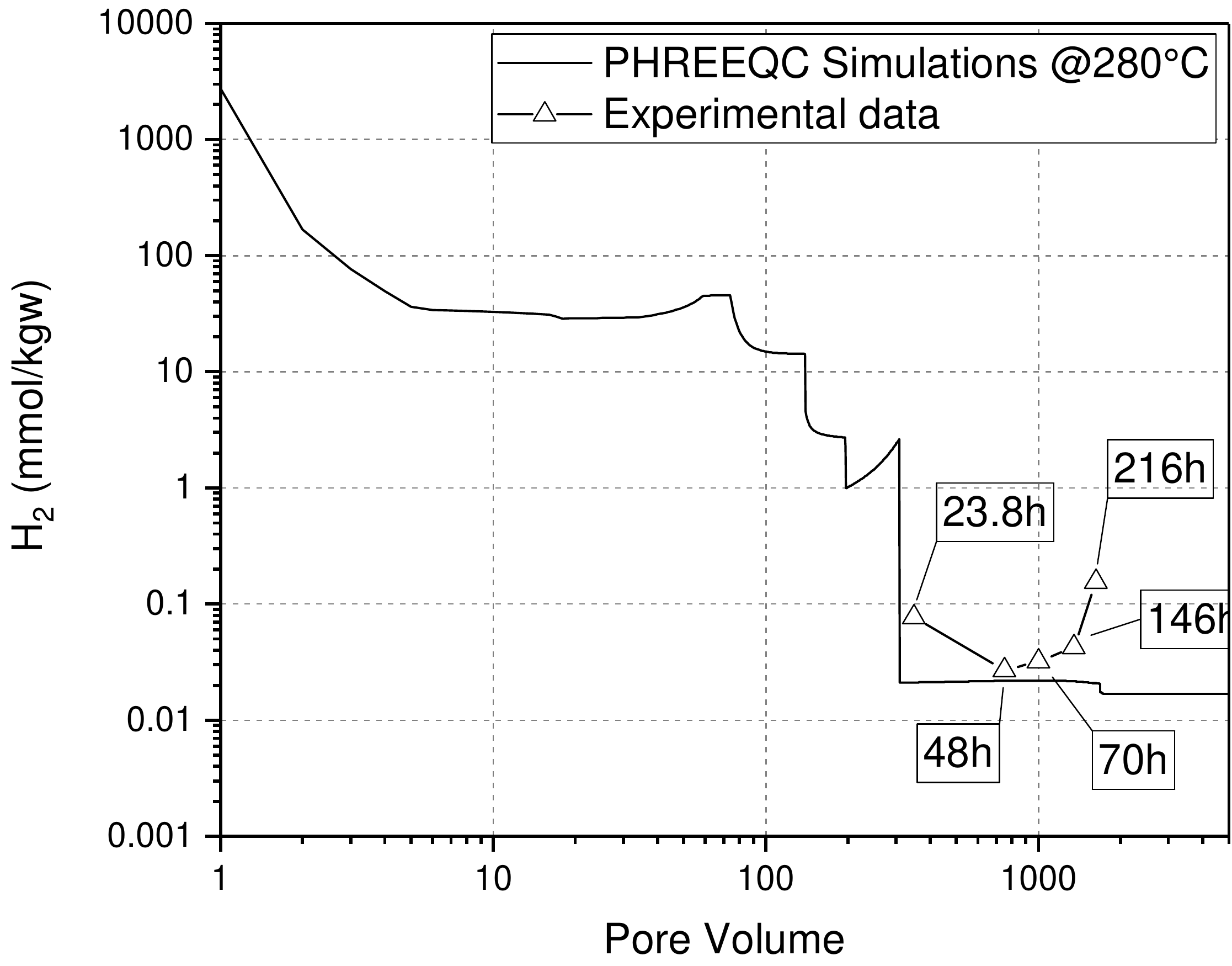}
	\caption{\label{fig:H2} Comparison between experiment (HTE - symbols) and simulations (solid line) in terms of \ce{H2} concentration. Labels on the experimental data points correspond to the time of sampling}
	\end{figure}

\section{Conclusion}

We presented in this article two experiments of reactive percolation of a \ce{NaHCO3}-rich brine at 160\textcelsius\ and 280\textcelsius\ in natural serpentinite cores in order to study the dynamic competition between serpentinization and carbonation of ultramafic formations either in natural or in engineered systems. During the 10 to 14 days of the experiments, the permeability of the cores dropped by several orders of magnitude, while the final yields at 280\textcelsius\ were 5.6\% for \ce{CO2} sequestration (quantity of \ce{CO2} stored over quantity injected) and 0.8\% for \ce{H2} production calculated from the total Fe content in the initial core. 

During the experiments, the cores clogged rapidly due to the precipitation of carbonates in the main percolation path, precipitation made possible by their fast reaction kinetics compared to the flow rate (Da $\gg$ 1). In contrast, precipitation of silicate minerals (serpentine) was found to be less impactful since they preferentially precipitates in low-flow zones due to their slow reaction kinetics. This discrepancy between high-flow zones and low-flow zones is directly linked to the pore size distribution of the starting protolith and justifies the use of natural cores for these kind of studies.

The control on the mineralogical assemblage by the relative kinetics of precipitation of the secondary phases and their relationship with advection, makes comparisons of natural samples with static batch experiments and numerical simulations complicated and potentially misleading. For example, such comparisons could lead to the underestimation of the quantity of fluid which percolated through the considered formation. This was for example the case in our experiments where the mineral assemblage was representative of an injected pore volume of 250 or less while the actual quantity of fluid injected in the cores was about ten times larger. 

In conclusion, results from the experiments highlight the importance of quantitatively analyzing the spatio-temporal lengthscales associated with the different chemical reactions occurring in the considered system and their relationship with the pore size distribution. These lengthscales are directly linked to the \Da\ number and to the type of reaction regime (transport-limited vs reaction-limited) allowing the interpretation of the final results in terms of a dynamic interplay between dissolution and precipitation, controlled by the local flow rate and the local pore geometry. In particular, the complex pore size distribution in natural rocks leads to very different behavior even for a homogeneous mineralogy. 

\section{Acknowledgements}
We would like to thank P. Penhoud for XRD and cathodoluminescence data. This research was supported by the LABEX Voltaire (ANR-10-LABX-100-01) and EQUIPEX PLANET (ANR-11-EQPX-0036). 

\section{Annexes}
	\subsection{Supplementary information 1: Material \& Methods }
	
	\subsubsection{Protocol for hydrogen sampling and quantification}
	A gas tight syringe was first filled with argon before being emptied and fixed to the sampling port. The syringe was then filled with the fluid  and the overpressure released in a beaker of deionized water. A small vial of 25 ml was then completely filled with deionized water, sealed with a septum and a volume of 10 ml of argon was injected through the septum in the vial to create a gas cap while the overpressure (the extra water) was released with an additional needle through the septum. Finally, the sampled fluid was injected in the vial while the overpressure was again being released. As at atmospheric pressure the solubility of hydrogen is negligible, this methodology ensures that the gas cap in the vial is a mixture of argon and hydrogen (plus any additional gas that may be dissolved in the sampled fluid), without any contamination from the atmosphere and with keeping the pressure in the vial at exactly the atmospheric pressure. The volume of argon also allows for replicates of the measurements and for most hydrogen samples, the measurement was performed at least 3 times. Each time, 1 ml of gas was sampled and immediately analyzed for hydrogen quantification by Gas Chromatography (PerkinElmer Clarus 580 - mounted with a Carboxen 1010PLOT Fused Silica capillary 30 m$\times$0.53 mm (ID)). Hydrogen was detected with a Thermal Conductivity Detector at 100\textcelsius. The carrier gas was nitrogen. Quantification of hydrogen was obtained from a calibration curve drawn with several known mixtures of Argon and hydrogen as described in \citep{Fauguerolles2016}. 
	
	\subsubsection{Micro-X-ray fluorescence}
Micro-X-ray fluorescence mapping were performed on sample sections using Bruker M4 Tornado spectrometer at Institut Terre et Environnement de Strasbourg (ITES, France). The X-ray tube consists of a Rh anode operating at 400 $\mu$A with an acceleration voltage of 50 kV. Polycapillary lenses were used to focus the X-ray beam down to 20 $\mu$m full-width-at-half-maximum at the sample surface. Two energy-dispersive X-ray detectors of 125 eV resolution each were used simultaneously to measure fluorescence spectra. Chemical maps were recorded on-the-fly with an acquisition time of 3 ms/pixel, a total of ~$\approx$40 cycles (dwell time $\approx$120 ms/pixel) and a 25 $\mu$m step interval in both directions. For each map, the scale bar corresponds to the intensity of K$\alpha$-lines of the elements (Mg, Al, Si, Ca, Cr, Fe, Ni) calculated from the integration of a specified region of interest (ROI) of the energy range of XRF spectra. Then, ROI maps (see electronic supplements, Figure XX) are used to calculate phase maps thanks to a Matlab-based code previously developed in Mu$\tilde{n}$oz et al. (2008) and Ulrich et al. (2014). The phase map calculation consists first of determining pure mineral phases that are expected to be present in the sample, in order to create standard spectra. Then, for each pixel of the map, a linear combination of the different standard spectra is performed to fit each single spectrum. Results provide quantitative phase maps showing the distribution of minerals previously identified in the sample (e.g., based on XRD and/or Raman analyses). This approach is particularly useful to highlight relationships between minerals, especially for the characterization of finely divided mineral assemblages, i.e., when the beam is larger than grain size (such as in this study).

\begin{figure}
	\centering \includegraphics[width=\textwidth]{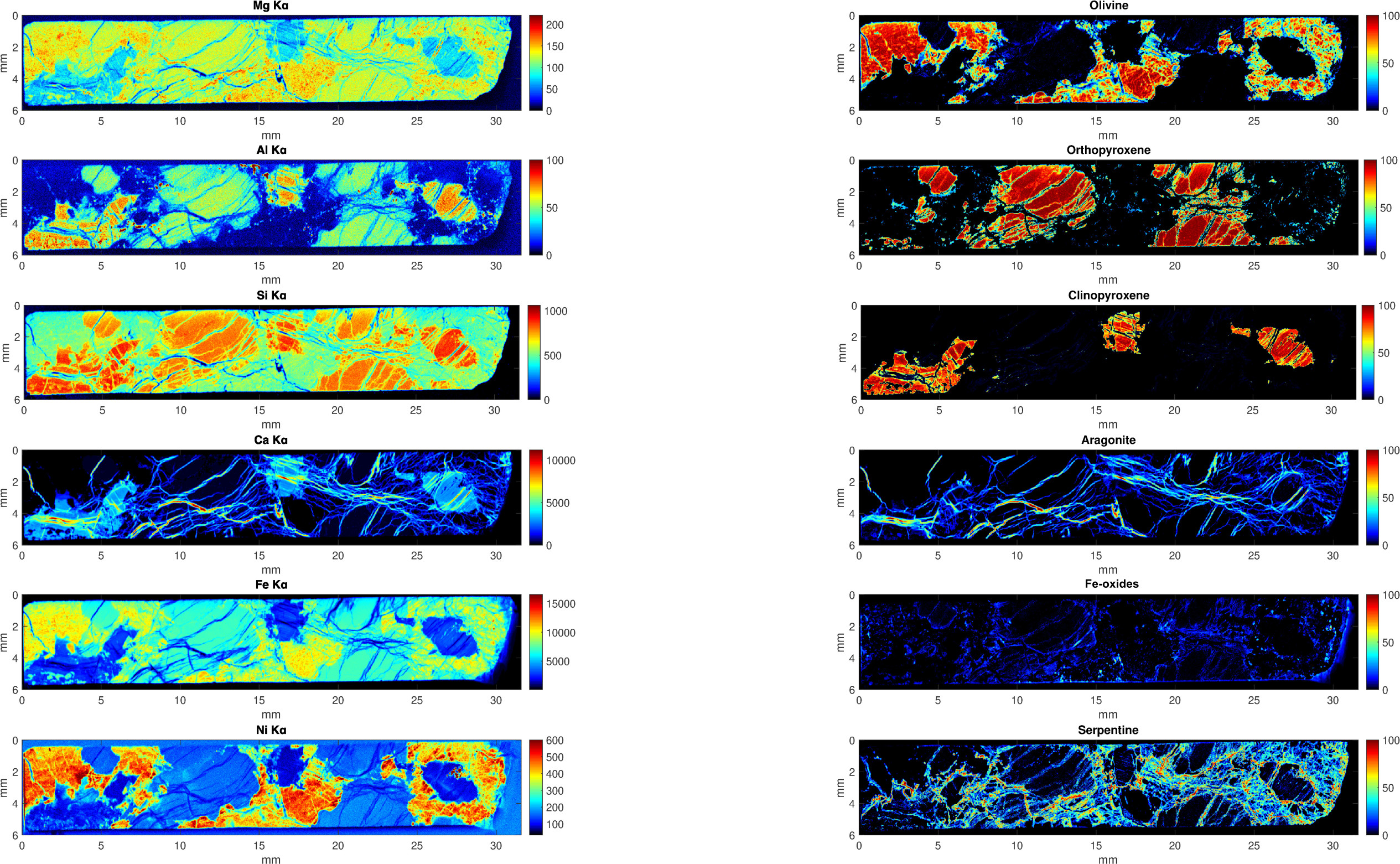}
	\caption{Elemental and phase reconstruction from XRF mapping for LTE}
\end{figure}

\begin{figure}
	\centering \includegraphics[width=\textwidth]{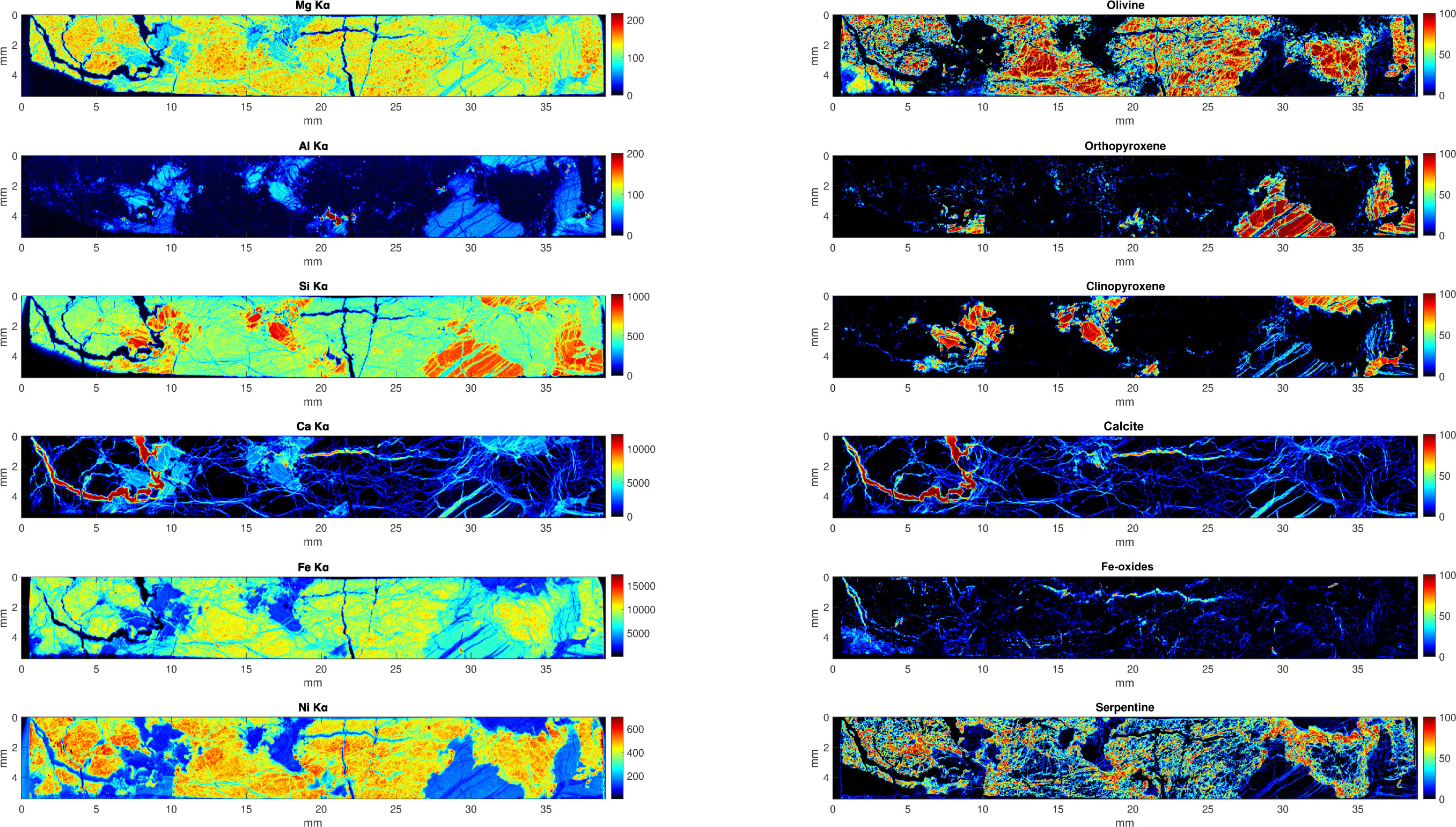}
	\caption{Elemental and phase reconstruction from XRF mapping for HTE}
\end{figure}

	\subsubsection{Raman Spectroscopy}
Raman spectra were obtained using a Renishaw InVIA Reflex microspectrometer (Institut des Sciences de la Terre d'Orléans [ISTO]-- Bureau de Recherches Géologiques et Minières [BRGM]). A laser (514 nm) was focused on the sample by a DM2500 Leica microscope equipped with a $\times$100 objective. Instrument control and Raman measurements were performed with the software package Renishaw Wire 4.0. Acquisition time was 1s,
and spectra were accumulated 20 times. The laser beam power at sample surface was set to $\approx$3 mW. 	

	\subsubsection{Cathodoluminescence}
Cathololuminescence maps were obtained with a Cathodyne under 15kV and 150$\mu$A conditions. Exposure time was adapted to obtain the best contrast between 2 and 8s.

\subsection{Supplementary Information 2: Different phases considered in the PHREEQC simulation and their composition}

\begin{table}
\centering 
\begin{small}

	\begin{tabular}{lll}
	\toprule[1.5pt]
	\multicolumn{3}{c}{\textbf{Starting minerals}} \\
	\midrule[1.5pt]
	\\
	\textbf{Phase} & \multicolumn{2}{l}{\textbf{Composition}} \\
	\midrule[1pt]
	Olivine & \multicolumn{2}{r}{$\ce{Mg_{1.8}Fe_{0.2}SiO4}$} \\
	\midrule
	Orthopyroxene & \multicolumn{2}{r}{$\ce{Mg_{0.84}Al_{0.12}Fe_{0.1}Si_{0.93}O3}$} \\
	\midrule 
	Clinopyroxene & \multicolumn{2}{r}{$\ce{Mg_{0.83}Fe_{0.09}Al_{0.31}Ca_{0.91}Si_{1.85}O6}$} \\
	\midrule
	Serpentine & \multicolumn{2}{r}{$\ce{Mg_{2.67}Fe_{0.27}Al_{0.04}Si_{1.89}O_{9.54}H_{5.8}}$} \\
	\midrule
	Aragonite & \multicolumn{2}{r}{$\ce{CaCO3}$} \\
	\midrule
	Magnetite & \multicolumn{2}{r}{$\ce{Fe3O4}$} \\
	\midrule[1.5pt]
	\\
	\\
	\midrule[1.5pt]
	\multicolumn{3}{c}{\textbf{Potential product minerals}} \\
	\midrule[1.5pt]
	\\
	\cmidrule[1.5pt]{1-1}
	\textbf{Solid Solutions} \\
	\cmidrule[1.5pt]{1-1}
	\\
	
	\textbf{Phase}	&	\multicolumn{2}{l}{\textbf{Composition}} \\
	\midrule[1pt]
		Brucite	&	\textbf{Brucite} \ce{Mg(OH)2}		& \textbf{Fe-brucite} \ce{Fe(OH)2}  \\
	\midrule
		Talc	&	\textbf{Mg-talc} \ce{Mg3Si4O10(OH)2} & \textbf{Minnesotaite} \ce{Fe3Si4O10(OH)2}  \\
	\midrule
		Serpentine & \textbf{Lizardite} \ce{Mg3Si2O5(OH)4}	& \textbf{Greenalite} $\ce{Fe(II)3Si2O5(OH)4}$  \\
					&  \textbf{Cronstedtite} $\ce{Fe(II)2Fe(III)2SiO5(OH)4}$ & \textbf{Amesite} \ce{Mg2Al(AlSiO5)(OH)4} \\
	\midrule
		Chlorite & \textbf{Clinochlore} \ce{Mg5Al(AlSi3)O10(OH)8} & \textbf{Chamosite} \ce{Fe5Al(AlSi3)O10(OH)8}  \\
	\midrule
		Tremolite & \textbf{Tremolite} \ce{(Ca2Mg5)Si8O22(OH)2} & \textbf{Ferrotremolite} \ce{(Ca2Fe5)Si8O22(OH)2}  \\
	\midrule
		Carbonate & \textbf{Calcite} \ce{CaCO3} & \textbf{Magnesite} \ce{MgCO3}  \\
				& \textbf{Siderite} \ce{FeCO3} \\
	\midrule[1pt]
	\\
	\cmidrule[1.5pt]{1-1}		
	\textbf{Pure Phases} \\
	\cmidrule[1.5pt]{1-1}
	\\
	\textbf{Phase} & \multicolumn{2}{r}{\textbf{Composition}} \\
	\midrule[1pt]
	Saponite(Na) 	&	\multicolumn{2}{r}{$\ce{Na_{0.33}Mg3Al_{0.33}Si_{3.67}O10(OH)2)}$} \\
	\midrule
	Vermiculite(Na) 	&	\multicolumn{2}{r}{$\ce{Na_{0.86}Mg3Si_{3.14}Al_{0.86}O10(OH)2})$} \\
	\midrule
	Hematite			&	\multicolumn{2}{r}{\ce{Fe2O3}} \\
	\midrule
	Quartz			&	\multicolumn{2}{r}{\ce{SiO2}} \\
	\bottomrule[1.5pt]
	\end{tabular}	
\end{small}
\end{table}

\subsection{Supplementary Information 3: EPMA analysis and whole rock composition}

\begin{table}[h]
\begin{scriptsize}
	\begin{tabular}{lcccccccccccccc}
	\toprule
	Mineral & \ce{Na2O} & \ce{MgO} & \ce{Al2O3} & \ce{SiO2} & \ce{K2O} & \ce{CaO} & \ce{TiO2} & \ce{MnO} & \ce{FeO} & \ce{Cr2O3} & \ce{NiO} & Total & Mg\# & Al/Si\\
	\midrule[1pt]
	\multicolumn{13}{c}{Starting Material} \\
	\midrule
	Olivine & 0.05 & 48.7 & 0.02 & 40.0 & 0.01 & 0.05 & 0.03 & 0.13 & 10.6 & 0.00 & 0.42 & 100 & 90 &  \\
	\midrule
	Orthopyroxene & 0.06 & 31.8 & 5.72 & 52.4 & 0.02 & 1.19 & 0.09 & 0.13 & 6.63 & 0.59 & 0.10 & 98.8 & 90 & 0.032 \\
	\midrule
	Clinopyroxene & 0.75	& 15.1 & 6.93 & 50.5 & 0.01 & 22.7 & 0.48 & 0.14 & 2.83 & 0.85 & 0.02 & 100.3 & 92 & 0.04 \\
	\midrule 
	Serpentine (mesh) & 0.02 & 37.7 & 0.2 & 39.4 & 0.01 & 0.07 & 0.04 & 0.07 & 6.24 & 0.02 & 0.33 & 84.0 & 91 & 0.0015 \\
	\midrule
	Serpentine (bastite) & 0.01 & 32.7 & 5.5 & 38.5 & 0.01 & 0.80 & 0.00 & 0.13 & 8.0 & 0.18 & 0.26 & 86.2 & 88 & 0.04 \\
	\midrule[1pt]
	\multicolumn{13}{c}{HTE} \\
	\midrule 
	Serpentine (mesh) & 0.13 & 37.1 & 0.53 & 39.1 & 0.04 & 0.37 & 0.02 & 0.01 & 7.26 & 0.00 & 0.31 & 84.9 & 90 & 0.004 \\
	\midrule
	Serpentine (bastite) & 0.31 & 35.2 & 5.02 & 40.1 & 0.08 & 0.11 & 0.05 & 0.01 & 4.54 & 0.86 & 0.17  & 86.5 & 93 & 0.037 \\ 
\bottomrule	
	\end{tabular}
\end{scriptsize}
\caption{\label{EPMAresults} Results from EPMA. Data were aquired at 15kV, 6na and 10s}
\end{table}

Mg\#\ is calculated as 2$\times$MgO/(MgO+Fe$_{tot}$+CaO) for clinopyroxene and MgO/(MgO+FeO) for the other minerals. The ratio Al/Si is calculated with atomic data.

\subsection{Supplementary Information 4: \Da\ numbers}

The \Da\ numbers were calculated as $Da = k/vc$ with $k$ the intrinsic reaction rate of precipitation or dissolution of the considered mineral, $v$ the Darcy velocity in the two percolating classes of pores (>62$\mu$m and [1-62]$\mu$m) and $c$ the solubility of the mineral in pure water at the pressure and temperature of the experiment. Due to the dearth of data on precipitation kinetics, the data for dissolution are being used for both precipitation and dissolution, except for magnesite whose precipitation behavior was studied by \citet{Saldi2012}. This obviously induces a error in the estimation Da, most likely an overestimation (usually precipitation is slower than dissolution for a given mineral). Another poorly constrained factor is the influence of dissolved \ce{CO2} on the precipitation/dissolution kinetics of silicates. According to \citet{Daval2013}, bicarbonate ions could enhance lizardite dissolution. On the contrary, in their review, \citet{Oelkers2018} conclude that dissolved \ce{CO2} does not impact directly olivine dissolution rate, and argue that any impact which could be measured in the literature is due to pH and ionic strength changes or to the formation of passivating Mg-bearing carbonates at the surface. In absence of any definitive data on this subject, the different calculated kinetics are considered as independent of \ce{CO2} concentration. Qualitatively, if \ce{CO2} had any effect, it would enhance dissolution rate and thus potentially reduce precipitation rates.

The velocity in each class of pores was calculated considering that each class consists in a bundle of cylindrical channels. For a cylindrical channel, the flow rate can be expressed as:

	\begin{equation}
		\label{Rh}
		Q = \frac{\Delta P}{Rh} \qquad \text{with} \qquad Rh = \frac{8 \eta L}{\pi r^4}
	\end{equation}
with $L$ the length of the channel, $r$ its radius. This means that for each class of pore, we can define a partial flow rate depending on the corresponding average pore radius, as well as the equivalent number of channels (i.e. pores). From Mercury Intrusion Porosimetry, we have the cumulative volumes for our 3 classes of pores of respectively 0.02ml/g, 0.002ml/g and 0.009ml/g (>62$\mu$m, [1-62]$\mu$m and <1$\mu$m). As a result, considering that the cylindrical channels span the whole length of the core, we have $N = V/(\pi r^2 L)$, and in terms of velocity, $v = Q/(N \pi R^2) = L/V \Delta P/Rh$. This allows to calculate roughly the ratio of the velocities in the different classes of pores such as $v_1 = 10^7 v_2 = 10^{16}v_3$. As expected, the main flow paths are the large fractures and the serpentine matrix is supporting almost no flow.

\begin{table}
\begin{footnotesize}
	\begin{tabular}{lrrlrrll}
	\toprule[1pt]
	\multirow{2}{*}{Mineral} & \multicolumn{2}{c}{160\textcelsius} & & \multicolumn{2}{c}{280\textcelsius} & & \multirow{2}{*}{References for reaction kinetics data}\\
	\cmidrule{2-3} \cmidrule{5-6}
			&	> 62 $\mu$m	&	[1-62]$\mu$m &  & > 62 $\mu$m	&	[1-62]$\mu$m & &\\
	\midrule[1pt]
	\\
	\textbf{Primary minerals} &	&	& & &	&	\\
	\cmidrule[1pt]{1-1}
	Olivine (Fo$_{90}$)	&	10$^{-8}$ 	& 100 	& &	1.5.10$^{-4}$		& 1500 & & \citep{Oelkers2008} \\
	\midrule
	Orthopyroxene (En$_{89}$) & 2.10$^{-5}$ & 	200	& &	 0.03	&	3.10$^{5}$ & & \citep{Palandri2004b}\\
	\midrule	
	Clinopyroxene (Di$_{91}$)	& 1.10$^{-9}$ &	0.01 & & 9.10$^{-8}$	& 0.9 & & \citep{Knauss1993} \\
	\midrule
	Aragonite 				& 3000	 & 3.10$^{10}$ & & 2.4.10$^4$ & 2.4.10$^{11}$ & & data for calcite \citep{Palandri2004b}\\
	\midrule[1pt]
	\\
	\textbf{Minerals potentially} & & & & & & & \\
	\textbf{precipitating or dissolving} & & & & & & &\\
	\cmidrule[1pt]{1-1}
		Serpentine				&	4.5.10$^{-6}$  & 45	&		&	 3.8.10$^{-4}$		&	3800 & &\citep{Orlando2011,Daval2013} \\
	\midrule
	Magnetite				&	0.05	&	5.5.10$^{5}$	&	& 0.03 	&	3.10$^{5}$ & &\citep{Palandri2004b} \\
	\midrule[1pt]
	\\
	\textbf{Precipitating minerals} & & & & & & & \\
	\cmidrule[1pt]{1-1}
	Calcite					&	4800 & 4.8.10$^{10}$ & & 4.5.10$^4$ & 45.10$^{11}$ & & \citep{Palandri2004b} \\
	\midrule
	Magnesite				&	2.10$^{-11}$			&	2.10$^{-4}$ 	& &	5.7.10$^{-11}$	&	5.7.10$^{-4}$ & & \citep{Saldi2012} \\
	\midrule
	Hematite 				&	40 & 40.10$^8$	&	&	3.5		& 3.5.10$^7$& & \citep{Palandri2004b} \\
	\midrule
		Talc						&	1.4.10$^{-6}$	&	14		&		&	1.1.10$^{-4}$	&	1100 & &\citep{Saldi2007}\\
	\midrule 
	Quartz					&	2.5.10$^{-6}$		&	25	&	& 2.10$^{-4}$	&	2000 & & \citep{Palandri2004b} \\
	\midrule
	Saponite					&	1.5.10$^{-6}$ & 15 & & 4.8.10$^{-5}$ & 48 & & \citep{Palandri2004b} \\
	\midrule[1pt]
	\\
	\textbf{Fluid velocity (m.s$^{-1}$)}	& 7.5.10$^{-5}$	&	7.5.10$^{-12}$ & & 5.2.10$^{-5}$ & 5.2.10$^{-12}$ & & \\
	\bottomrule[1pt]

	\end{tabular}
	\caption{\label{tab:Da} \Da\ numbers for the different minerals in the experiments. The first rows correspond to the primary minerals which only dissolve. Serpentine and magnetite are already present in the composition and can either dissolve or precipitate (middle rows) and finally, the bottom rows lists minerals that can only precipitate.}
	\end{footnotesize}
\end{table}

\bibliographystyle{gca}
\bibliography{biblio}

\begin{thebibliography}{80}
\providecommand{\natexlab}[1]{#1}
\expandafter\ifx\csname urlstyle\endcsname\relax
  \providecommand{\doi}[1]{doi:\discretionary{}{}{}#1}\else
  \providecommand{\doi}{doi:\discretionary{}{}{}\begingroup
  \urlstyle{rm}\Url}\fi

\bibitem[{Andreani et~al.(2009)Andreani, Luquot, Gouze, Godard,
  Hois{\{}$\backslash$'e{\}} and Gibert}]{Andreani2009}
Andreani M., Luquot L., Gouze P., Godard M., Hois{\{}$\backslash$'e{\}} E. and
  Gibert B. (2009) {Experimental Study of Carbon Sequestration Reactions
  Controlled by the Percolation of CO 2 -Rich Brine through Peridotites}.
\newblock \emph{Environ. Sci. {\&} Technol.}, \textbf{43(4)},
  1226--1231.
\newblock \doi{10.1021/es8018429}.

\bibitem[{Andreani and M{\'{e}}nez(2019)}]{Andreani2019}
Andreani M. and M{\'{e}}nez B. (2019) \emph{{New perspectives on abiotic
  organic synthesis and processing during hydrothermal alteration of the
  oceanic lithosphere}}.

\bibitem[{Appelo(2015)}]{Appelo2015}
Appelo C.A. (2015) {Principles, caveats and improvements in databases for
  calculating hydrogeochemical reactions in saline waters from 0 to 200°C and
  1 to 1000atm}.
\newblock \emph{Appl. Geochem.}, \textbf{55}, 62--71.
\newblock \doi{10.1016/j.apgeochem.2014.11.007}.

\bibitem[{Beinlich et~al.(2012)Beinlich, Pl{\"{u}}mper, H{\"{o}}velmann,
  Austrheim and Jamtveit}]{Beinlich2012}
Beinlich A., Pl{\"{u}}mper O., H{\"{o}}velmann J., Austrheim H. and Jamtveit B.
  (2012) {Massive serpentinite carbonation at Linnajavri, N-Norway}.
\newblock \emph{Terra Nova}, \textbf{24(6)}, 446--455.
\newblock \doi{10.1111/j.1365-3121.2012.01083.x}.

\bibitem[{Blanc et~al.(2012)Blanc, Lassin, Piantone, Azaroual, Jacquemet,
  Fabbri and Gaucher}]{Blanc2012}
Blanc P., Lassin A., Piantone P., Azaroual M., Jacquemet N., Fabbri A. and
  Gaucher E.C. (2012) {Thermoddem: A geochemical database focused on low
  temperature water/rock interactions and waste materials}.
\newblock \emph{Appl. Geochem.}, \textbf{27(10)}, 2107--2116.
\newblock \doi{10.1016/j.apgeochem.2012.06.002}.

\bibitem[{Bonatti et~al.(1980)Bonatti, Lawrence, Hamlyn and
  Breger}]{Bonatti1980}
Bonatti E., Lawrence J.R., Hamlyn P.R. and Breger D. (1980) {Aragonite from
  deep sea ultramafic rocks}.
\newblock \emph{Geochim. Cosmochim. Acta}, \textbf{44(8)}, 1207--1214.
\newblock \doi{10.1016/0016-7037(80)90074-5}.

\bibitem[{Boschi et~al.(2009)Boschi, Dini, Dallai, Ruggieri and
  Gianelli}]{Boschi2009}
Boschi C., Dini A., Dallai L., Ruggieri G. and Gianelli G. (2009) {Enhanced
  CO2-mineral sequestration by cyclic hydraulic fracturing and Si-rich fluid
  infiltration into serpentinites at Malentrata (Tuscany, Italy)}.
\newblock \emph{Chem. Geol.}, \textbf{265(1-2)}, 209--226.
\newblock \doi{10.1016/j.chemgeo.2009.03.016}.

\bibitem[{Cannat et~al.(2010)Cannat, Fontaine and Escart{\'{i}}n}]{Cannat2010}
Cannat M., Fontaine F. and Escart{\'{i}}n J. (2010) {Serpentinization and
  associated hydrogen and methane fluxes at slow spreading ridges}.
\newblock pages 241--264.
\newblock \doi{10.1029/2008GM000760}.

\bibitem[{Chavagnac et~al.(2013)Chavagnac, Ceuleneer, Monnin, Lansac, Hoareau
  and Boulart}]{Chavagnac2013}
Chavagnac V., Ceuleneer G., Monnin C., Lansac B., Hoareau G. and Boulart C.
  (2013) {Mineralogical assemblages forming at hyperalkaline warm springs
  hosted on ultramafic rocks: A case study of Oman and Ligurian ophiolites}.
\newblock \emph{Geochem. Geophys. Geosyst.}, \textbf{14(7)},
  2474--2495.
\newblock \doi{10.1002/ggge.20146}.

\bibitem[{Cipolli et~al.(2004)Cipolli, Gambardella, Marini, Ottonello and
  Zuccolini}]{Cipolli2004}
Cipolli F., Gambardella B., Marini L., Ottonello G. and Zuccolini M.V. (2004)
  {Geochemistry of high-pH waters from serpentinites of the Gruppo di Voltri
  (Genova, Italy) and reaction path modeling of CO2sequestration in
  serpentinite aquifers}.
\newblock \emph{Appl. Geochem.}, \textbf{19(5)}, 787--802.
\newblock \doi{10.1016/j.apgeochem.2003.10.007}.

\bibitem[{Escario et~al.(2018)Escario, Godard, Gouze and
  Leprovost}]{Escario2018}
Escario S., Godard M., Gouze P. and Leprovost R. (2018) {Experimental study of
  the effects of solute transport on reaction paths during incipient
  serpentinization}.
\newblock \emph{Lithos}, \textbf{323}, 191--207.
\newblock \doi{10.1016/j.lithos.2018.09.020}.

\bibitem[{Escart{\'{i}}n et~al.(1997)Escart{\'{i}}n, Hirth and
  Evans}]{Escartin1997}
Escart{\'{i}}n J., Hirth G. and Evans B. (1997) {Effects of serpentinization on
  the lithospheric strength and the style of normal faulting at slow-spreading
  ridges}.
\newblock \emph{Earth Planet. Sci. Lett.}, \textbf{151(3-4)},
  181--189.
\newblock \doi{10.1016/S0012-821X(97)81847-X}.

\bibitem[{Falk and Kelemen(2015)}]{Falk2015}
Falk E.S. and Kelemen P.B. (2015) {Geochemistry and petrology of listvenite in
  the Samail ophiolite, Sultanate of Oman: Complete carbonation of peridotite
  during ophiolite emplacement}.
\newblock \emph{Geochim. Cosmochim. Acta}, \textbf{160}, 70--90.
\newblock \doi{10.1016/j.gca.2015.03.014}.

\bibitem[{Farough et~al.(2016)Farough, Moore, Lockner and Lowell}]{Farough2016}
Farough A., Moore D.E., Lockner D.A. and Lowell R.P. (2016) {Evolution of
  fracture permeability of ultramafic rocks undergoing serpentinization at
  hydrothermal conditions: An experimental study}.
\newblock \emph{Geochem. Geophys. Geosyst.}, \textbf{17(1)}, 44--55.
\newblock \doi{10.1002/2015GC005973}.

\bibitem[{Fr{\"{u}}h-Green et~al.(2004)Fr{\"{u}}h-Green, Connolly, Plas, Kelley
  and Grob{\'{e}}ty}]{Fruh-Green2004}
Fr{\"{u}}h-Green G.L., Connolly J.A., Plas A., Kelley D.S. and Grob{\'{e}}ty B.
  (2004) {Serpentinization of oceanic peridotites: Implications for geochemical
  cycles and biological activity}.
\newblock \emph{Geoph. Monog. Series}, \textbf{144(November 2014)},
  119--136.
\newblock \doi{10.1029/144GM08}.

\bibitem[{Gadikota et~al.(2020)Gadikota, Matter, Kelemen, Brady and
  Park}]{Gadikota2020}
Gadikota G., Matter J., Kelemen P., Brady P.V. and Park A.H.A. (2020)
  {Elucidating the differences in the carbon mineralization behaviors of
  calcium and magnesium bearing alumino-silicates and magnesium silicates for
  CO2 storage}.
\newblock \emph{Fuel}, \textbf{277(March)}, 117900.
\newblock \doi{10.1016/j.fuel.2020.117900}.

\bibitem[{Gaillardet et~al.(1999)Gaillardet, Dupr{\'{e}}, Louvat and
  All{\`{e}}gre}]{Gaillardet1999}
Gaillardet J., Dupr{\'{e}} B., Louvat P. and All{\`{e}}gre C.J. (1999) {Global
  silicate weathering and CO2 consumption rates deduced from the chemistry of
  large rivers}.
\newblock \emph{Chem. Geol.}, \textbf{159(1-4)}, 3--30.
\newblock \doi{10.1016/S0009-2541(99)00031-5}.

\bibitem[{G{\'{i}}slason et~al.(2018)G{\'{i}}slason, Sigurdard{\'{o}}ttir,
  Arad{\'{o}}ttir and Oelkers}]{Gislason2018}
G{\'{i}}slason S.R., Sigurdard{\'{o}}ttir H., Arad{\'{o}}ttir E.S. and Oelkers
  E.H. (2018) {A brief history of CarbFix: Challenges and victories of the
  project's pilot phase}.
\newblock In \emph{Enrgy Proced.}, volume 146.
\newblock \doi{10.1016/j.egypro.2018.07.014}.

\bibitem[{Godard et~al.(2013)Godard, Luquot, Andreani and Gouze}]{Godard2013}
Godard M., Luquot L., Andreani M. and Gouze P. (2013) {Incipient hydration of
  mantle lithosphere at ridges: A reactive-percolation experiment}.
\newblock \emph{Earth Plan. Sci. Lett.}, \textbf{371-372},
  92--102.
\newblock \doi{10.1016/j.epsl.2013.03.052}.

\bibitem[{Grozeva et~al.(2017)Grozeva, Klein, Seewald and Sylva}]{Grozeva2017}
Grozeva N.G., Klein F., Seewald J.S. and Sylva S.P. (2017) {Experimental study
  of carbonate formation in oceanic peridotite}.
\newblock \emph{Geochim. et Cosmochim. Acta}, \textbf{199}, 264--286.
\newblock \doi{10.1016/j.gca.2016.10.052}.

\bibitem[{Hansen et~al.(2005)Hansen, Dipple, Gordon and Kellett}]{Hansen2005}
Hansen L.D., Dipple G.M., Gordon T.M. and Kellett D.A. (2005) {Carbonated
  serpentinite (listwanite) at Atlin, British Columbia: A geological analogue
  to carbon dioxide sequestration}.
\newblock \emph{Can. Mineral.}, \textbf{43(1)}, 225--239.
\newblock \doi{10.2113/gscanmin.43.1.225}.

\bibitem[{Hatakeyama et~al.(2017)Hatakeyama, Katayama, Hirauchi and
  Michibayashi}]{Hatakeyama2017}
Hatakeyama K., Katayama I., Hirauchi K.I. and Michibayashi K. (2017) {Mantle
  hydration along outer-rise faults inferred from serpentinite permeability}.
\newblock \emph{Sci. Rep.-UK}, \textbf{7(1)}, 1--8.
\newblock \doi{10.1038/s41598-017-14309-9}.

\bibitem[{Hinsken et~al.(2017)Hinsken, Br{\"{o}}cker, Strauss and
  Bulle}]{Hinsken2017}
Hinsken T., Br{\"{o}}cker M., Strauss H. and Bulle F. (2017) {Geochemical,
  isotopic and geochronological characterization of listvenite from the Upper
  Unit on Tinos, Cyclades, Greece}.
\newblock \emph{Lithos}, \textbf{282-283}, 281--297.
\newblock \doi{10.1016/j.lithos.2017.02.019}.

\bibitem[{H{\"{o}}velmann et~al.(2011)H{\"{o}}velmann, Austrheim, Beinlich and
  {Anne Munz}}]{Hovelmann2011}
H{\"{o}}velmann J., Austrheim H., Beinlich A. and {Anne Munz} I. (2011)
  {Experimental study of the carbonation of partially serpentinized and
  weathered peridotites}.
\newblock \emph{Geochim. Cosmochim. Acta}, \textbf{75(22)}, 6760--6779.
\newblock \doi{10.1016/j.gca.2011.08.032}.

\bibitem[{H{\"{o}}velmann et~al.(2012)H{\"{o}}velmann, Austrheim and
  Jamtveit}]{Hovelmann2012}
H{\"{o}}velmann J., Austrheim H. and Jamtveit B. (2012) {Microstructure and
  porosity evolution during experimental carbonation of a natural peridotite}.
\newblock \emph{Chem. Geol.}, \textbf{334}, 254--265.
\newblock \doi{10.1016/j.chemgeo.2012.10.025}.

\bibitem[{Hybler et~al.(2016)Hybler, Sejkora and Vencl{\'{i}}k}]{Hybler2016}
Hybler J., Sejkora J. and Vencl{\'{i}}k V. (2016) {Polytypism of cronstedtite
  from Pohled, Czech Republic}.
\newblock \emph{Eur. J. Mineral.}, \textbf{28(4)}, 765--775.
\newblock \doi{10.1127/ejm/2016/0028-2532}.

\bibitem[{{International Energy Agency}(2019)}]{InternationalEnergyAgency2019}
{International Energy Agency} (2019) {The Future of Hydrogen}.
\newblock Technical report, International Energy Agency.

\bibitem[{Iyer et~al.(2008)Iyer, Jamtveit, Mathiesen, Malthe-S{\o}renssen and
  Feder}]{Iyer2008}
Iyer K., Jamtveit B., Mathiesen J., Malthe-S{\o}renssen A. and Feder J. (2008)
  {Reaction-assisted hierarchical fracturing during serpentinization}.
\newblock \emph{Earth Plan. Sci. Lett.}, \textbf{267(3-4)},
  503--516.
\newblock \doi{10.1016/j.epsl.2007.11.060}.

\bibitem[{Jamtveit et~al.(2000)Jamtveit, Austrheim and
  Malthe-Sorenssen}]{Jamtveit2000}
Jamtveit B., Austrheim H. and Malthe-Sorenssen A. (2000) {Accelerated hydration
  of the Earth's deep crust induced by stress perturbations}.
\newblock \emph{Nature}, \textbf{408(6808)}, 75--78.
\newblock \doi{10.1038/35040537}.

\bibitem[{Kelemen et~al.(2018)Kelemen, Aines, Bennett, Benson, Carter, Coggon,
  de~Obeso, Evans, Gadikota, Dipple, Godard, Harris, Higgins, Johnson, Kourim,
  Lafay, Lambart, Manning, Matter, Michibayashi, Morishita, No{\"{e}}l,
  Okazaki, Renforth, Robinson, Savage, Skarbek, Spiegelman, Takazawa, Teagle,
  Urai and Wilcox}]{Kelemen2018}
Kelemen P., Aines R., Bennett E., Benson S., Carter E., Coggon J., de~Obeso J.,
  Evans O., Gadikota G., Dipple G., Godard M., Harris M., Higgins J., Johnson
  K., Kourim F., Lafay R., Lambart S., Manning C., Matter J., Michibayashi K.,
  Morishita T., No{\"{e}}l J., Okazaki K., Renforth P., Robinson B., Savage H.,
  Skarbek R., Spiegelman M., Takazawa E., Teagle D., Urai J. and Wilcox J.
  (2018) {In situ carbon mineralization in ultramafic rocks: Natural processes
  and possible engineered methods}.
\newblock \emph{Enrgy Proced.}, \textbf{146(August)}, 92--102.
\newblock \doi{10.1016/j.egypro.2018.07.013}.

\bibitem[{Kelemen and Hirth(2012)}]{Kelemen2012}
Kelemen P.B. and Hirth G. (2012) {Reaction-driven cracking during retrograde
  metamorphism: Olivine hydration and carbonation}.
\newblock \emph{Earth Plan. Sci. Lett.}, \textbf{345-348}, 81--89.
\newblock \doi{10.1016/j.epsl.2012.06.018}.

\bibitem[{Kelemen et~al.(2011)Kelemen, Matter, Streit, Rudge, Curry and
  Blusztajn}]{Kelemen2011}
Kelemen P.B., Matter J., Streit E.E., Rudge J.F., Curry W.B. and Blusztajn J.
  (2011) {Rates and Mechanisms of Mineral Carbonation in Peridotite: Natural
  Processes and Recipes for Enhanced, in situ CO 2 Capture and Storage}.
\newblock \emph{Annu. Rev. Earth Pl. Sc.}, \textbf{39(1)},
  545--576.
\newblock \doi{10.1146/annurev-earth-092010-152509}.

\bibitem[{Kelemen and Matter(2008)}]{Kelemen2008}
Kelemen P.B. and Matter J.M. (2008) {In situ carbonation of peridotite for CO2
  storage}.
\newblock \emph{P. Nat. Acad. Sci.},
  \textbf{105(45)}, 17295--17300.
\newblock \doi{10.1073/pnas.0805794105}.

\bibitem[{Klein et~al.(2013)Klein, Bach and McCollom}]{Klein2013a}
Klein F., Bach W. and McCollom T.M. (2013) {Compositional controls on hydrogen
  generation during serpentinization of ultramafic rocks}.
\newblock \emph{Lithos}, \textbf{178}, 55--69.
\newblock \doi{10.1016/j.lithos.2013.03.008}.

\bibitem[{Klein and Garrido(2011)}]{Klein2011}
Klein F. and Garrido C.J. (2011) {Thermodynamic constraints on mineral
  carbonation of serpentinized peridotite}.
\newblock \emph{Lithos}, \textbf{126(3-4)}, 147--160.
\newblock \doi{10.1016/j.lithos.2011.07.020}.

\bibitem[{Klein and McCollom(2013)}]{Klein2013}
Klein F. and McCollom T.M. (2013) {From serpentinization to carbonation: New
  insights from a CO2 injection experiment}.
\newblock \emph{Earth Plan. Sci. Lett.}, \textbf{379(31)},
  137--145.
\newblock \doi{10.1016/j.epsl.2013.08.017}.

\bibitem[{Kondratiuk et~al.(2017)Kondratiuk, Tredak, Upadhyay, Ladd and
  Szymczak}]{Kondratiuk2017}
Kondratiuk P., Tredak H., Upadhyay V., Ladd A.J. and Szymczak P. (2017)
  {Instabilities and finger formation in replacement fronts driven by an
  oversaturated solution}.
\newblock \emph{J. Geophys. Res-Sol. Ea.}, \textbf{122(8)},
  5972--5991.
\newblock \doi{10.1002/2017JB014169}.

\bibitem[{Lafay et~al.(2018)Lafay, Montes-Hernandez, Renard and
  Vonlanthen}]{Lafay2018}
Lafay R., Montes-Hernandez G., Renard F. and Vonlanthen P. (2018)
  {Intracrystalline Reaction-Induced Cracking in Olivine Evidenced by Hydration
  and Carbonation Experiments}.
\newblock \emph{Minerals}, \textbf{8(9)}, 412.
\newblock \doi{10.3390/min8090412}.

\bibitem[{Luhmann et~al.(2017{\natexlab{a}})Luhmann, Tutolo, Bagley, Mildner,
  Scheuermann, Feinberg, Ignatyev and Seyfried}]{Luhmann2017b}
Luhmann A.J., Tutolo B.M., Bagley B.C., Mildner D.F., Scheuermann P.P.,
  Feinberg J.M., Ignatyev K. and Seyfried W.E. (2017{\natexlab{a}}) {Chemical
  and physical changes during seawater flow through intact dunite cores: An
  experimental study at 150–200 °C}.
\newblock \emph{Geochim. et Cosmochim. Acta}, \textbf{214}, 86--114.
\newblock \doi{10.1016/j.gca.2017.07.020}.

\bibitem[{Luhmann et~al.(2017{\natexlab{b}})Luhmann, Tutolo, Bagley, Mildner,
  Seyfried and Saar}]{Luhmann2017}
Luhmann A.J., Tutolo B.M., Bagley B.C., Mildner D.F.R., Seyfried W.E. and Saar
  M.O. (2017{\natexlab{b}}) {Permeability, porosity, and mineral surface area
  changes in basalt cores induced by reactive transport of CO 2 -rich brine}.
\newblock \emph{Water Resour. Res.}, \textbf{53(3)}, 1908--1927.
\newblock \doi{10.1002/2016WR019216}.

\bibitem[{Malvoisin(2015)}]{Malvoisin2015}
Malvoisin B. (2015) {Mass transfer in the oceanic lithosphere: Serpentinization
  is not isochemical}.
\newblock \emph{Earth Plan. Sci. Lett.}, \textbf{430}, 75--85.
\newblock \doi{10.1016/j.epsl.2015.07.043}.

\bibitem[{Malvoisin et~al.(2017)Malvoisin, Brantut and
  Kaczmarek}]{Malvoisin2017}
Malvoisin B., Brantut N. and Kaczmarek M.A. (2017) {Control of serpentinisation
  rate by reaction-induced cracking}.
\newblock \emph{Earth Plan. Sci. Lett.}, \textbf{476}, 143--152.
\newblock \doi{10.1016/j.epsl.2017.07.042}.

\bibitem[{Malvoisin et~al.(2012)Malvoisin, Brunet, Carlut, Roum{\'{e}}jon and
  Cannat}]{Malvoisin2012}
Malvoisin B., Brunet F., Carlut J., Roum{\'{e}}jon S. and Cannat M. (2012)
  {Serpentinization of oceanic peridotites: 2. Kinetics and processes of San
  Carlos olivine hydrothermal alteration}.
\newblock \emph{J. Geophys. Res-Sol. Ea.}, \textbf{117(4)}.
\newblock \doi{10.1029/2011JB008842}.

\bibitem[{Marcaillou et~al.(2011)Marcaillou, Mu{\~{n}}oz, Vidal, Parra and
  Harfouche}]{Marcaillou2011}
Marcaillou C., Mu{\~{n}}oz M., Vidal O., Parra T. and Harfouche M. (2011)
  {Mineralogical evidence for H2 degassing during serpentinization at
  300°C/300bar}.
\newblock \emph{Earth Plan. Sci. Lett.}, \textbf{303(3-4)},
  281--290.
\newblock \doi{10.1016/j.epsl.2011.01.006}.

\bibitem[{Martin and Fyfe(1970)}]{Martin1970}
Martin B. and Fyfe W. (1970) {Some experimental and theoretical observations on
  the kinetics of hydration reactions with particular reference to
  serpentinization}.
\newblock \emph{Chem. Geol.}, \textbf{6(9)}, 185--202.
\newblock \doi{10.1016/0009-2541(70)90018-5}.

\bibitem[{McCollom and Bach(2009)}]{Mccollom2009}
McCollom T.M. and Bach W. (2009) {Thermodynamic constraints on hydrogen
  generation during serpentinization of ultramafic rocks}.
\newblock \emph{Geochim. Cosmochim. Acta}, \textbf{73(3)}, 856--875.
\newblock \doi{10.1016/j.gca.2008.10.032}.

\bibitem[{McCollom et~al.(2020)McCollom, Klein, Moskowitz, Berqu{\'{o}}, Bach
  and Templeton}]{McCollom2020}
McCollom T.M., Klein F., Moskowitz B., Berqu{\'{o}} T.S., Bach W. and Templeton
  A.S. (2020) {Hydrogen generation and iron partitioning during experimental
  serpentinization of an olivine–pyroxene mixture}.
\newblock \emph{Geochim. Cosmochim. Acta}, \textbf{282}, 55--75.
\newblock \doi{10.1016/j.gca.2020.05.016}.

\bibitem[{McCollom and Seewald(2001)}]{Mccollom2001}
McCollom T.M. and Seewald J.S. (2001) {A reassessment of the potential for
  reduction of dissolved CO2 to hydrocarbons during serpentinization of
  olivine}.
\newblock \emph{Geochim. et Cosmochim. Acta}, \textbf{65(21)}, 3769--3778.
\newblock \doi{10.1016/S0016-7037(01)00655-X}.

\bibitem[{Mcgrail et~al.(2006)Mcgrail, Schaef, Ho, Chien, Dooley and
  Davidson}]{McGrail2006}
Mcgrail B.P., Schaef H.T., Ho A.M., Chien Y.j., Dooley J.J. and Davidson C.L.
  (2006) {Potential for carbon dioxide sequestration in flood basalts}.
\newblock \textbf{111(August)}, 1--13.
\newblock \doi{10.1029/2005JB004169}.

\bibitem[{Menzel et~al.(2018)Menzel, Garrido, {L{\'{o}}pez
  S{\'{a}}nchez-Vizca{\'{i}}no}, Marchesi, Hidas, Escayola and {Delgado
  Huertas}}]{Menzel2018}
Menzel M.D., Garrido C.J., {L{\'{o}}pez S{\'{a}}nchez-Vizca{\'{i}}no} V.,
  Marchesi C., Hidas K., Escayola M.P. and {Delgado Huertas} A. (2018)
  {Carbonation of mantle peridotite by CO 2 -rich fluids: the formation of
  listvenites in the Advocate ophiolite complex (Newfoundland, Canada)}.
\newblock \emph{Lithos}, \textbf{323}.
\newblock \doi{10.1016/j.lithos.2018.06.001}.

\bibitem[{Miller et~al.(2017)Miller, Mayhew, Ellison, Kelemen, Kubo and
  Templeton}]{Miller2017}
Miller H.M., Mayhew L.E., Ellison E.T., Kelemen P., Kubo M. and Templeton A.S.
  (2017) {Low temperature hydrogen production during experimental hydration of
  partially-serpentinized dunite}.
\newblock \emph{Geochim. et Cosmochim. Acta}, \textbf{209}, 161--183.
\newblock \doi{10.1016/j.gca.2017.04.022}.

\bibitem[{Morgan et~al.(2015)Morgan, Wilson, Madsen, Gozukara and
  Habsuda}]{Morgan2015}
Morgan B., Wilson S.A., Madsen I.C., Gozukara Y.M. and Habsuda J. (2015)
  {Increased thermal stability of nesquehonite (MgCO 3 {\textperiodcentered}3H
  2 O) in the presence of humidity and CO 2 : Implications for low-temperature
  CO 2 storage}.
\newblock \emph{Int. J. Greenh. Gas Con.}, \textbf{39},
  366--376.
\newblock \doi{10.1016/j.ijggc.2015.05.033}.

\bibitem[{Morrow et~al.(2001)Morrow, Moore and Lockner}]{Morrow2001}
Morrow C.A., Moore D.E. and Lockner D.A. (2001) {Permeability reduction in
  granite under hydrothermal conditions the granite decreased with time t ,
  following the exponential relation portional to temperature and ranged
  between between Figure 1 . Cylindrical sample configurations for permeability
  tests}.
\newblock \emph{J. Geophys. Res.}, \textbf{106(Table 1)},
  30,551--30,560.

\bibitem[{Muller et~al.(2009)Muller, Qi, Mackie, Pruess and Blunt}]{Muller2009}
Muller N., Qi R., Mackie E., Pruess K. and Blunt M.J. (2009) {CO2 injection
  impairment due to halite precipitation}.
\newblock \emph{Enrgy Proced.}, \textbf{1(1)}, 3507--3514.
\newblock \doi{10.1016/j.egypro.2009.02.143}.

\bibitem[{Neubeck et~al.(2011)Neubeck, Duc, Bastviken, Crill and
  Holm}]{Neubeck2011}
Neubeck A., Duc N.T., Bastviken D., Crill P. and Holm N.G. (2011) {Formation of
  H2 and CH4by weathering of olivine at temperatures between 30 and 70°C}.
\newblock \emph{Geochem. T.}, \textbf{12(1)}, 6.
\newblock \doi{10.1186/1467-4866-12-6}.

\bibitem[{O'Connor et~al.(2005)O'Connor, Dahlin, Rush, Gerdemann, Penner and
  Nilsen}]{Oconnor2005}
O'Connor W., Dahlin D.C., Rush G., Gerdemann S.J., Penner L. and Nilsen D.
  (2005) {Aqueous Mineral Carbonation: Mineral Availability, Pretreatment,
  Reaction Parametrics, and Process Studies}.
\newblock \emph{Doe/Arc-Tr-04-002}, \textbf{(April)}, 1--19.
\newblock \doi{10.13140/RG.2.2.23658.31684}.

\bibitem[{Oelkers et~al.(2008)Oelkers, Gislason and Matter}]{Oelkers2008}
Oelkers E.H., Gislason S.R. and Matter J. (2008) {Mineral carbonation of CO2}.
\newblock \emph{Elements}, \textbf{4(5)}, 333--337.
\newblock \doi{10.2113/gselements.4.5.333}.

\bibitem[{Osselin et~al.(2015{\natexlab{a}})Osselin, Fabbri, Fen-Chong, Dangla,
  Pereira and Lassin}]{Osselin2015b}
Osselin F., Fabbri A., Fen-Chong T., Dangla P., Pereira J.M. and Lassin A.
  (2015{\natexlab{a}}) {Stress from NaCl crystallisation by carbon dioxide
  injection in aquifers}.
\newblock \emph{Environmental Geotechnics}, \textbf{2(5)}, 280--291.
\newblock \doi{10.1680/envgeo.13.00057}.

\bibitem[{Osselin et~al.(2015{\natexlab{b}})Osselin, Fabbri, Fen-Chong,
  Pereira, Lassin and Dangla}]{Osselin2015}
Osselin F., Fabbri A., Fen-Chong T., Pereira J.M., Lassin A. and Dangla P.
  (2015{\natexlab{b}}) {Experimental investigation of the influence of
  supercritical state on the relative permeability of Vosges sandstone}.
\newblock \emph{CR - Mecanique}, \textbf{343(9)}.
\newblock \doi{10.1016/j.crme.2015.06.009}.

\bibitem[{Osselin et~al.(2013)Osselin, Fen-Chong, Fabbri, Lassin, Pereira and
  Dangla}]{Osselin2013a}
Osselin F., Fen-Chong T., Fabbri A., Lassin A., Pereira J.M. and Dangla P.
  (2013) {Dependence on injection temperature and on aquifer's petrophysical
  properties of the local stress applying on the pore wall of a crystallized
  pore in the context of
  CO{\textless}inf{\textgreater}2{\textless}/inf{\textgreater} storage in deep
  saline aquifers}.
\newblock \emph{Eur. Phys. J-Appl. Phys.}, \textbf{64(2)}, 21101.
\newblock \doi{10.1051/epjap/2013120529}.

\bibitem[{Osselin et~al.(2016)Osselin, Kondratiuk, Budek, Cybulski, Garstecki
  and Szymczak}]{Osselin2016}
Osselin F., Kondratiuk P., Budek A., Cybulski O., Garstecki P. and Szymczak P.
  (2016) {Microfluidic observation of the onset of reactive infiltration
  instability in an analog fracture}.
\newblock \emph{Geophys. Res. Lett.}, \textbf{43(13)}.
\newblock \doi{10.1002/2016GL069261}.

\bibitem[{Osselin et~al.(2019)Osselin, Kondratiuk, Cybulski, Garstecki and
  Szymczak}]{Osselin2019a}
Osselin F., Kondratiuk P., Cybulski O., Garstecki P. and Szymczak P. (2019)
  {Microfluidic measurement of the dissolution rate of gypsum in water using
  the reactive infiltration-instability}.
\newblock \emph{E3S Web Conf.}, \textbf{98}, 04010.
\newblock \doi{10.1051/e3sconf/20199804010}.

\bibitem[{Paulick et~al.(2006)Paulick, Bach, Godard, {De Hoog}, Suhr and
  Harvey}]{Paulick2006}
Paulick H., Bach W., Godard M., {De Hoog} J.C., Suhr G. and Harvey J. (2006)
  {Geochemistry of abyssal peridotites (Mid-Atlantic Ridge, 15°20'N, ODP Leg
  209): Implications for fluid/rock interaction in slow spreading
  environments}.
\newblock \emph{Chem. Geol.}, \textbf{234(3-4)}, 179--210.
\newblock \doi{10.1016/j.chemgeo.2006.04.011}.

\bibitem[{Peuble et~al.(2015{\natexlab{a}})Peuble, Andreani, Godard, Gouze,
  Barou, {Van De Moortele}, Mainprice and Reynard}]{Peuble2015}
Peuble S., Andreani M., Godard M., Gouze P., Barou F., {Van De Moortele} B.,
  Mainprice D. and Reynard B. (2015{\natexlab{a}}) {Carbonate mineralization in
  percolated olivine aggregates: Linking effects of crystallographic
  orientation and fluid flow}.
\newblock \emph{Am. Mineral.}, \textbf{100(2-3)}, 474--482.
\newblock \doi{10.2138/am-2015-4913}.

\bibitem[{Peuble et~al.(2018)Peuble, Andreani, Gouze, Pollet-Villard, Reynard
  and {Van de Moortele}}]{Peuble2018}
Peuble S., Andreani M., Gouze P., Pollet-Villard M., Reynard B. and {Van de
  Moortele} B. (2018) {Multi-scale characterization of the incipient
  carbonation of peridotite}.
\newblock \emph{Chem. Geol.}, \textbf{476(May 2017)}, 150--160.
\newblock \doi{10.1016/j.chemgeo.2017.11.013}.

\bibitem[{Peuble et~al.(2019)Peuble, Godard, Gouze, Leprovost, Martinez and
  Shilobreeva}]{Peuble2019}
Peuble S., Godard M., Gouze P., Leprovost R., Martinez I. and Shilobreeva S.
  (2019) {Control of CO 2 on flow and reaction paths in olivine-dominated
  basements: An experimental study}.
\newblock \emph{Geochim. Cosmochim. Acta}, \textbf{252}, 16--38.
\newblock \doi{10.1016/j.gca.2019.02.007}.

\bibitem[{Peuble et~al.(2015{\natexlab{b}})Peuble, Godard, Luquot, Andreani,
  Martinez and Gouze}]{Peuble2015a}
Peuble S., Godard M., Luquot L., Andreani M., Martinez I. and Gouze P.
  (2015{\natexlab{b}}) {CO2 geological storage in olivine rich basaltic
  aquifers: New insights from reactive-percolation experiments}.
\newblock \emph{Appl. Geochem.}, \textbf{52}, 174--190.
\newblock \doi{10.1016/j.apgeochem.2014.11.024}.

\bibitem[{Power et~al.(2013)Power, Harrison, Dipple, Wilson, Kelemen, Hitch and
  Southam}]{Power2013a}
Power I.M., Harrison A.L., Dipple G.M., Wilson S.A., Kelemen P.B., Hitch M. and
  Southam G. (2013) {Carbon Mineralization: From Natural Analogues to
  Engineered Systems}.
\newblock \emph{Reviews in Mineralogy and Geochemistry}, \textbf{77(1)},
  305--360.
\newblock \doi{10.2138/rmg.2013.77.9}.

\bibitem[{Putnis(2009)}]{Putnis2009}
Putnis A. (2009) {Mineral Replacement Reactions}.
\newblock \emph{Rev. Mineral. Geochem.}, \textbf{70(1)},
  87--124.
\newblock \doi{10.2138/rmg.2009.70.3}.

\bibitem[{Reynard(2013)}]{Reynard2013}
Reynard B. (2013) {Serpentine in active subduction zones}.
\newblock \emph{Lithos}, \textbf{178}, 171--185.
\newblock \doi{10.1016/j.lithos.2012.10.012}.

\bibitem[{Roum{\'{e}}jon et~al.(2014)Roum{\'{e}}jon, Cannat, Agrinier, Godard
  and Andreani}]{Roumejon2014}
Roum{\'{e}}jon S., Cannat M., Agrinier P., Godard M. and Andreani M. (2014)
  {Serpentinization and fluid pathways in tectonically exhumed peridotites from
  the southwest Indian ridge (62-65°E)}.
\newblock \emph{J. Petrol.}, \textbf{56(4)}, 703--734.
\newblock \doi{10.1093/petrology/egv014}.

\bibitem[{Ruiz-Agudo et~al.(2014)Ruiz-Agudo, Putnis and
  Putnis}]{Ruiz-Agudo2014}
Ruiz-Agudo E., Putnis C.V. and Putnis A. (2014) {Coupled dissolution and
  precipitation at mineral-fluid interfaces}.
\newblock \emph{Chem. Geol.}, \textbf{383}, 132-146
\newblock \doi{10.1016/j.chemgeo.2014.06.007}.

\bibitem[{Saldi et~al.(2012)Saldi, Schott, Pokrovsky, Gautier and
  Oelkers}]{Saldi2012}
Saldi G.D., Schott J., Pokrovsky O.S., Gautier Q. and Oelkers E.H. (2012) {An
  experimental study of magnesite precipitation rates at neutral to alkaline
  conditions and 100-200°C as a function of pH, aqueous solution composition
  and chemical affinity}.
\newblock \emph{Geochim. Cosmochim. Acta}, \textbf{83}, 93--109.
\newblock \doi{10.1016/j.gca.2011.12.005}.

\bibitem[{Seifritz(1990)}]{Seifritz1990}
Seifritz W. (1990) {CO2 disposal by means of silicates}.
\newblock \emph{Nature}, \textbf{345(6275)}, 486--486.
\newblock \doi{10.1038/345486b0}.

\bibitem[{Tominaga et~al.(2017)Tominaga, Beinlich, Lima, Tivey, Hampton, Weiss
  and Harigane}]{Tominaga2017}
Tominaga M., Beinlich A., Lima E.A., Tivey M.A., Hampton B.A., Weiss B. and
  Harigane Y. (2017) {Multi-scale magnetic mapping of serpentinite
  carbonation}.
\newblock \emph{Nat. Com.}, \textbf{8(1)}.
\newblock \doi{10.1038/s41467-017-01610-4}.

\bibitem[{Tutolo et~al.(2018)Tutolo, Luhmann, Tosca and Seyfried}]{Tutolo2018}
Tutolo B.M., Luhmann A.J., Tosca N.J. and Seyfried W.E. (2018)
  {Serpentinization as a reactive transport process: The brucite silicification
  reaction}.
\newblock \emph{Earth Plan. Sci. Lett.}, \textbf{484}, 385--395.
\newblock \doi{10.1016/j.epsl.2017.12.029}.

\bibitem[{Ulrich et~al.(2014)Ulrich, Mu{\~{n}}oz, Guillot, Cathelineau, Picard,
  Quesnel, Boulvais and Couteau}]{Ulrich2014}
Ulrich M., Mu{\~{n}}oz M., Guillot S., Cathelineau M., Picard C., Quesnel B.,
  Boulvais P. and Couteau C. (2014) {Dissolution-precipitation processes
  governing the carbonation and silicification of the serpentinite sole of the
  New Caledonia ophiolite}.
\newblock \emph{Contributions to Mineralogy and Petrology}, \textbf{167(1)},
  1--19.
\newblock \doi{10.1007/s00410-013-0952-8}.

\bibitem[{Voigt et~al.(2018)Voigt, Marieni, Clark, G{\'{i}}slason and
  Oelkers}]{Voigt2018}
Voigt M., Marieni C., Clark D.E., G{\'{i}}slason S.R. and Oelkers E.H. (2018)
  {Evaluation and refinement of thermodynamic databases for mineral
  carbonation}.
\newblock \emph{Enrgy Proced.}, \textbf{146}, 81--91.
\newblock \doi{10.1016/j.egypro.2018.07.012}.

\bibitem[{Wang et~al.(2019)Wang, Watanabe, Okamoto, Nakamura and
  Komai}]{Wang2019}
Wang J., Watanabe N., Okamoto A., Nakamura K. and Komai T. (2019) {Pyroxene
  control of H2 production and carbon storage during water-peridotite-CO2
  hydrothermal reactions}.
\newblock \emph{Int. J. Hydrogen Energ.}, \textbf{44(49)},
  26835--26847.
\newblock \doi{10.1016/j.ijhydene.2019.08.161}.

\bibitem[{Zhang and Dawe(2000)}]{Zhang2000}
Zhang Y. and Dawe R.A. (2000) {Influence of Mg2+ on the kinetics of calcite
  precipitation and calcite crystal morphology}.
\newblock \emph{Chem. Geol.}, \textbf{163(1-4)}, 129--138.
\newblock \doi{10.1016/S0009-2541(99)00097-2}.

\end{thebibliography}

\end{document}